\documentclass[12pt]{article}
\usepackage{epstopdf,graphicx,epsfig,rotating}
\usepackage{rotating,datetime}
\usepackage{times}
\usepackage{epstopdf, graphicx}
\usepackage{amssymb,amsmath}
\usepackage[superscript]{cite}
\usepackage[normalem]{ulem}
\usepackage{color}
\usepackage[labelfont=bf]{caption}

\def\be{\begin{equation}}
\def\ee{\end{equation}}
\def\bea{\begin{eqnarray}}
\def\eea{\end{eqnarray}}

\newenvironment{sciabstract}{%
\begin{quote} \bf}
{\end{quote}}

\def \kms            {${\rm km~s}^{-1}$}

\def \lsun {L$_\odot$}

\newcommand{\spt}{SPT2349-56}
\newcommand{\mgas}{M$_{\rm gas}$}
\newcommand{\mstar}{M_{\star}}

\newcommand{\msun}{M$_{\odot}$}
\newcommand{\mpy}{\msun\,yr$^{-1}$}
\newcommand{\ssubmm}{$S_{\rm 870\,\mu m}$}
\newcommand{\smm}{S_{\rm 1.1\,mm}}

\newcounter{lastnote}

\title{\bf A massive core for a cluster of galaxies at a redshift of 4.3 }
\author{
	\parbox{\textwidth}{  
		\normalsize T.\ B.\ Miller$^{1,2}$, S.\ C.\ Chapman$^{1,3,4}$, M.\ Aravena$^5$, M.\ L.\ N.\ Ashby$^6$, C.\ C.\ Hayward$^{6,7}$, J.\ D.\ Vieira$^8$, A.\ Wei\ss$^9$, %End of main authors
		A.\ Babul$^{10}$,  M.\ B\'ethermin$^{11}$, C.\ M.\ Bradford$^{12,13}$, M.\ Brodwin$^{14}$, J. E. Carlstrom$^{15,16,17,18}$, Chian-Chou Chen$^{19}$, D.\ J.\ M.\ Cunningham$^{1,20}$, C.\ De Breuck $^{19}$, A.\ H.\ Gonzalez$^{21}$, T.\ R.\ Greve$^{22}$,J. Harnett$^{23}$, Y.\ Hezaveh$^{24}$, K.\ Lacaille$^{1,25}$, K.\ C.\ Litke$^{26}$, J.\ Ma$^{21}$, M.\ Malkan$^{27}$, D.\ P.\ Marrone$^{26}$, W.\ Morningstar$^{24}$, E.\ J.\ Murphy$^{28}$, D.\ Narayanan$^{21}$, E.\ Pass$^{1,29}$, R.\ Perry$^{1}$, K.\ A.\ Phadke$^8$, D.\ Rennehan$^{10}$, K.\ M.\ Rotermund$^1$, J.\ Simpson$^{30,31}$, J.\ S.\ Spilker$^{26}$, J.\ Sreevani$^8$, A.\ A.\ Stark$^{6}$, M.\ L.\ Strandet$^{9,32}$ \& A.\ L.\ Strom$^{33}$}
	\vspace*{6pt} \\
	\parbox{\textwidth}{ \scriptsize
		$^1$Department of Physics and Atmospheric Science, Dalhousie University, Halifax, Canada\\
		$^2$Department of Astronomy, Yale University, 52 Hillhouse Avenue, New Haven, CT, 06511, USA\\
		$^3$Institute of Astronomy, Madingley Road, Cambridge CB3 0HA, UK \\
		$^4$National Research Council, Herzberg Astronomy \& Astrophysics, 5071 West Saanich Road, Victoria, BC, V9E 2E7, Canada\\
		$^5$N\'ucleo de Astronom\'ia, Facultad de Ingenier\'ia y Ciencias, Universidad Diego Portales, Av. Ejercito 441, Santiago, Chile\\
		$^6$Harvard-Smithsonian Center for Astrophysics, 60 Garden Street, Cambridge, MA 02138, USA \\
		$^7$Center for Computational Astrophysics, Flatiron Institute, 162 Fifth Avenue, New York, NY 10010, USA\\
		$^8$Department of Astronomy, University of Illinois, 1002 West Green Street, Urbana, IL 61801, USA \\
		$^9$Max-Planck-Institut fur Radioastronomie, Auf dem H\"ugel 69 D-53121 Bonn, Germany\\
		$^{10}$Department of Physics and Astronomy, University of Victoria, Victoria, BC V8P 1A1, Canada \\
		$^{11}$Aix-Marseille Universit\'e, CNRS, LAM, Laboratoire d'Astrophysique de Marseille, Marseille, France\\
		$^{12}$California Institute of Technology, 1200 E. California Boulevard, Pasadena, CA 91125, USA\\
		$^{13}$Jet Propulsion Laboratory, 4800 Oak Grove Drive, Pasadena, CA 91109, USA\\
		$^{14}$Department of Physics and Astronomy, University of Missouri, 5110 Rockhill Road, Kansas City, MO 64110, USA\\
		$^{15}$ Kavli Institute for Cosmological Physics, University of Chicago, 5640 South Ellis Avenue, Chicago, IL 60637, USA\\ 
		$^{16}$Department of Physics, University of Chicago, 5640 South Ellis Avenue, Chicago, IL 60637, USA\\
		$^{17}$Enrico Fermi Institute, University of Chicago, 5640 South Ellis Avenue, Chicago, IL 60637, USA\\
		$^{18}$Department of Astronomy and Astrophysics, University of Chicago, 5640 South Ellis Avenue, Chicago, IL 60637, USA\\
		$^{19}$European Southern Observatory, Karl Schwarzschild Stra\ss e 2, 85748 Garching, Germany\\
		$^{20}$Department of Astronomy and Physics, Saint Mary's University, Halifax, Nova Scotia, Canada \\
		$^{21}$Department of Astronomy, University of Florida, Bryant Space Sciences Center, Gainesville, FL 32611, USA\\
		$^{22}$Department of Physics and Astronomy, University College London, Gower Street, London WC1E 6BT, UK\\
		$^{23}$School of Physics, University of Sydney, Sydney, NSW 2006, Australia\\
		$^{24}$Kavli Institute for Particle Astrophysics and Cosmology, Stanford University, Stanford, CA 94305, USA\\
		$^{25}$Department of Physics and Astronomy, McMaster University, Hamilton, ON L8S4M1, Canada\\
		$^{26}$Steward Observatory, University of Arizona, 933 North Cherry Avenue, Tucson, AZ 85721, USA\\
		$^{27}$Department of Physics and Astronomy, University of California, Los Angeles, CA 90095, USA \\
		$^{28}$National Radio Astronomy Observatory, 520 Edgemont Road, Charlottesville, VA 22903, USA\\
		$^{29}$Department of Physics and Astronomy, University of Waterloo, 200 University Avenue West, Waterloo, ON N2L 3G1, Canada\\
		$^{30}$Institute for Astronomy, University of Edinburgh, Royal Observatory, Blackford Hill, Edinburgh EH9 3HJ, UK \\
		$^{31}$Centre for Extragalactic Astronomy, Department of Physics, Durham University, South Road, Durham DH1 3LE, UK\\
		$^{32}$International Max Planck Research School (IMPRS) for Astronomy and Astrophysics, Universities of Bonn and Cologne, Germany \\
		$^{33}$Observatories of The Carnegie Institution for Science, 813 Santa Barbara Street, Pasadena, CA 91101, USA\\
	}
	\date{}
}

\begin{document}

\maketitle

\begin{sciabstract}
Massive galaxy clusters are now  found as early as \textbf{$\sim$ 3} billion years after the Big Bang,  
 containing stars that formed at even earlier epochs \cite{Wang2016,Mantz2017,Stanford2012}. 
The high-redshift progenitors of these galaxy clusters, termed `protoclusters', are identified in cosmological simulations with the highest dark matter overdensities.\cite{Springel2005a,Overzier2009a,Chiang2017}
While their observational signatures are less well defined compared to virialized clusters with a substantial hot intra-cluster medium (ICM), protoclusters are expected to contain extremely massive galaxies that can be observed as luminous starbursts\cite{Miley2008}. 
Recent claimed detections of protoclusters hosting such starbursts\cite{Casey2015,Chapman2009,Tamura2009,Ma2015}
do not support the kind of rapid cluster core formation expected in simulations\cite{Chiang2013} because these  structures  contain only a handful of starbursting galaxies spread throughout a broad structure, with poor evidence for  eventual collapse into a protocluster.
Here we report  that the source \spt\ consists of at least 14 gas-rich galaxies all lying at \textbf{\textit z = 4.31} based on sensitive observations 
  of carbon monoxide and ionized carbon.  We demonstrate that each of these galaxies is forming stars 
between 50 and 1000 times faster than our own Milky Way, and all are located within a projected region only $\sim$ 130 kiloparsecs in diameter. This galaxy surface density is more than 10 times the average blank field value (integrated over all redshifts) and $>$1000 times the average field volume density.  
The velocity dispersion ($\sim$ 410 km s$^{-1}$) of these galaxies and enormous gas and star formation densities suggest that this system represents a galaxy cluster core at an advanced stage of formation 
when the Universe was only 1.4 billion years old. A comparison with other known protoclusters at high redshifts shows that \spt\  is a uniquely massive and dense system that could be building one of the most massive structures  in the Universe today.  
\end{sciabstract}

\begin{figure*}
	\centering
 	\includegraphics[width= 0.495\textwidth]{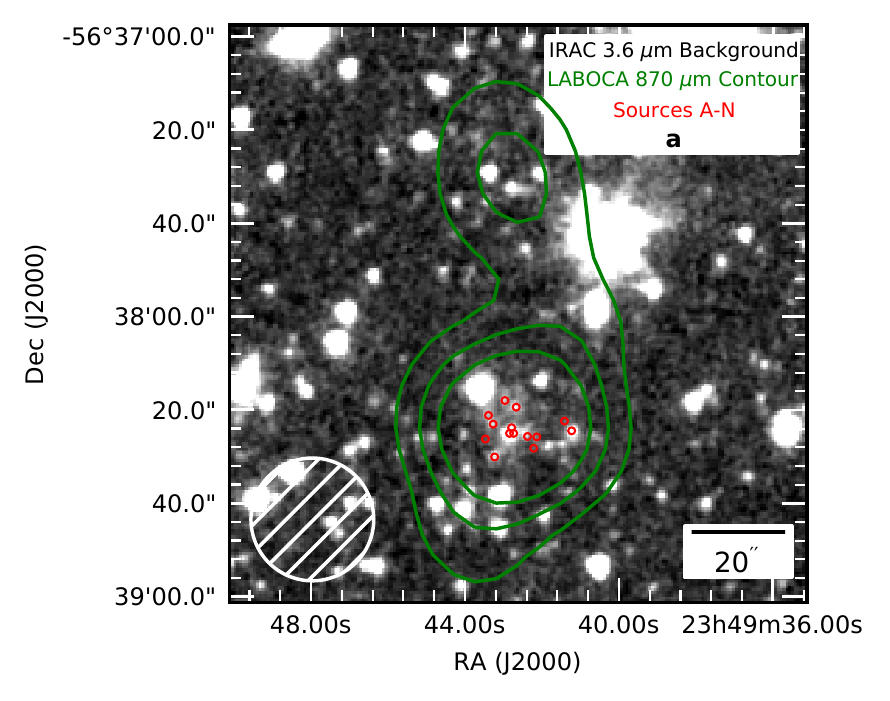}
 	\includegraphics[width= 0.495\textwidth]{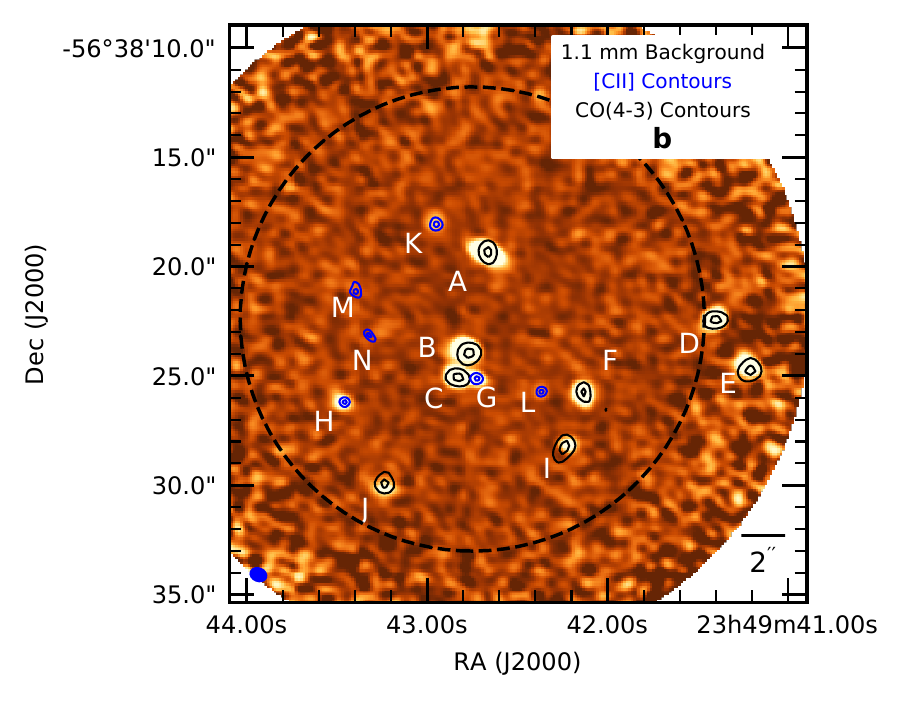}
 	\includegraphics[width= 1.02\textwidth]{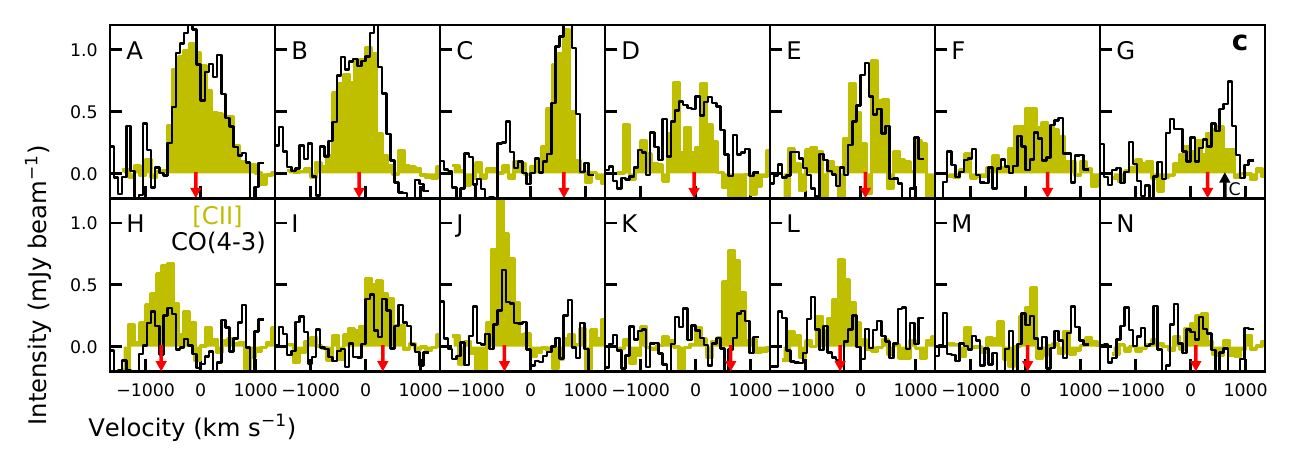}
 	
	\caption{\textbf{The SPT2349-56 field and spectra of the constituent galaxies.} 
	(a) The LABOCA 870-$\mu$m contours of \spt\ overlaid on the IRAC 3.6-$\mu$m  image; the 26$''$  beam at 870\,$\mu$m  is shown. Contours represent SNR = 3,7 and 9. The small circles show the locations of the 14 protocluster sources.	
	(b) ALMA band 7 imaging (276\,GHz, 1.1\,mm) displaying the 14 confirmed protocluster sources, labeled A-N. Black (blue) contours denote the points 75\% and 90\% of the peak flux  for each  source from the CO(4-3) ([CII]). The dashed black line shows where the primary beam is 50\% the maximum. 
	The filled blue ellipse shows the 0.4$''$ naturally weighted synthesized beam. 
	(c) CO(4-3) spectra (black lines) and [CII] spectra overlaid (shaded yellow bars)  for all 14 sources centred at the biweight cluster redshift $z = 4.304$.  
	The [CII] spectra are scaled down in flux by a factor of ten for clarity of presentation. The red arrows show the velocity offsets determined by fitting a Gaussian profile to the CO(4-3) spectra for all sources except for G, H, K, L, M, and N, for which we used [CII] because these are not detected in CO(4-3). 
}
	\label{fig:sources}
\end{figure*}

In a multi-band survey over 2500 deg$^2$ of sky, the South Pole Telescope (SPT) discovered a population of rare ($n\sim0.04$ deg$^{-2}$), extremely bright ($S_{\rm 1.4\,mm} > 20$\,mJy) millimeter-selected sources \cite{Vieira2010,Mocanu2013}. The Atacama Large Millimeter Array (ALMA) 870-$\mu$m %
imaging showed that more than $90\%$ of these SPT-selected sources are single high-redshift submillimeter galaxies (SMGs)\cite{Vieira2013} with intrinsic flux densities of $S_{\rm 870\,\mu m} = 5-10$\,mJy, but gravitationally lensed by factors of $5-20$ \cite{Spilker2016}, with a median redshift $z\sim4$ \cite{Strandet2016}. 
However, $\sim10$\% of these sources show no evidence for lensing and may instead be intrinsically very luminous galaxies or even groups of multiple rapidly star-forming galaxies.
The brightest such source in the SPT 2500 deg$^2$ survey, \spt\ ($S_{\rm 1.4\,mm}$=23.3\,mJy), 
is revealed by LABOCA (a low resolution bolometer camera on the APEX telescope)
 observations at 870\,$\mu$m to consist of  two elongated sources with a combined flux density \ssubmm\ $\sim110$\,mJy (Fig.~\ref{fig:sources}),  
with the brighter southern source comprising $\sim77$\,mJy of this flux density.
An ALMA redshift survey\cite{Strandet2016} further resolved \spt\ into a pair of bright 3-mm sources associated with the southern LABOCA source, with both lying at $z = 4.3$. 
To better understand the nature of this structure, deep ALMA spectral imaging of the brighter southern peak of the extended LABOCA source was undertaken. A 358-GHz map containing the redshifted [CII]$_{\rm 1900.5\,GHz}$ line was used to search for line-emitting galaxies. 
A blind spectral line survey (described in the Methods) was performed on the data cube, revealing 14 $z\sim4.31$ line emitters at high significance (SNR $>$7). Twelve of these emitters are individually detected in the 1.1-mm continuum map at $>5\sigma$, with 1.1-mm flux densities ranging from 0.2-5\,mJy (Fig.~\ref{fig:sources}). The remaining two line emitters (M,N) are both detected at lower significance  in the 1.1-mm continuum map but have robust IRAC infrared counterparts (Extended Data Table~1, Extended Data Fig.~\ref{fig:IRAC}). Eight of these sources are also detected ($>5\sigma$) in the CO(4-3) line.
The ALMA spectra are shown in Fig.~\ref{fig:sources}.

The measurements of both the continuum and spectral lines of the 14 galaxies allow us to estimate their star formation rates (SFRs) and gas masses (Tables~1 \& Extended Data Table 1). The physical properties of these sources indicate that this protocluster already harbors massive galaxies that are rapidly forming stars from an abundant gas supply. The two brightest sources, A \& B, have SFRs in excess of 1000 solar masses per year (\mpy) within their resolved $\sim3$-kpc radii. The total SFR of the 14 sources is $6000 \pm 600$ \mpy.  Multi-colour imaging with Herschel-SPIRE (250, 350, 500-$\mu$m), in addition to the 870-$\mu$m LABOCA map, show that the northern LABOCA structure is also consistent with lying at $z = 4.3$ (see Methods). The sources detected in the ALMA 870-$\mu$m imaging therefore comprise just 50\% of the total flux density of the southern LABOCA source and 36\% of the total LABOCA flux density, suggesting that the $\sim500$ kpc extent of the protocluster contains 16,500 \mpy of star formation. Modelling the spectral energy distribution based on this combined submillimeter photometry yields an IR luminosity (from 8-1100\,$\mu$m) of $(8.0\pm1.0)\times10^{13}$ times the solar luminosity (L$_\odot$). The gas masses of the 14 protocluster galaxies, estimated from CO(4-3), or [CII] if undetected in CO(4-3) (see Methods), range from  $1\times10^{10}$ to $1\times10^{11}$ \msun, with a total gas mass of $\sim6\times10^{11}$  $\left({\rm X_{CO}}/0.8\right)$ \msun.  
A follow-up survey of colder molecular gas in CO(2-1) with the ATCA radio telescope detects the bulk of this large gas repository,  especially in the central region near sources B, C, \& G, and confirms that  the assumed line intensity ratio, CO(4-3) to CO(1-0), used in the Methods when calculating the total gas mass is consistent with the average measurements from ATCA.

The detected ALMA sources also enables an initial estimate of the mass of the protocluster.
We determine the mean redshift using the biweight estimator\cite{Beers1990} 
as $\left<z\right>_{\rm bi} = 4.3040^{+0.0020}_{-0.0019}$.
The velocity dispersion of the galaxy distribution is $\sigma_{\rm bi} = 408^{+82}_{-56}$ \kms\ according to the biweight method\cite{Beers1990}, which is the standard approach for galaxy samples of this size. Other common methods (gapper\cite{Beers1990}, Gaussian fit) agree to within 3\% and provide similar errors.   
Under the assumption that \spt\ is approximately virialized, the mass-dispersion
relation for galaxy clusters\cite{Evrard2008} indicates a dynamical mass of $M_{\rm
dyn} = (1.16 \pm 0.70) \times 10^{13}$ M$_\odot$, which is an upper limit if the system has not yet virialized.
Given the possible selection effect of requiring a bright source (S$_{\rm 1.4mm}$ $>$ 15 mJy) within the 1$'$ SPT beam for detection, we also further consider the possibility that our structure may represent 
an end-on filament
being projected into a compact but unbound configuration,
rather than a single bound halo. Our analysis in the Methods suggests this is not as likely as a relatively bound system in a massive halo, given the velocity dispersion measured as a function of position, and other supporting arguments. However we cannot rule this possibility out completely, and further analysis and observations of the larger angular scale of the structure will be required to more fully understand the nature of this system.

If the total halo mass represented by these 14 SMGs is indeed $\sim10^{13}$ \msun, then the
protocluster is a viable progenitor of a $>10^{15}$ \msun\ galaxy cluster comparable to the Coma cluster at $z = 0$ (Fig.~\ref{fig:comparison}), which we deduce by comparison to simulations that track the hierarchical halo growth of a $\sim10^{15}$ \msun\ galaxy cluster from early epochs\cite{Chiang2013}.
The location of \spt\ in this plane suggests a very massive descendant, but we caution that N-body simulations indicate that it is difficult to reliably predict $z = 0$ halo mass from the halo mass at a given epoch due to the large halo-to-halo variation in dark matter halo growth histories \cite{Cole2008}.

To study the relative overdensity and concentration of \spt, it is desirable to compare with
other active protoclusters at high redshift.  \spt\ is highly overdense, as 
it harbors 10 SMGs with $\smm \gtrsim 0.5$ (a level at which we are complete, with uniform sensitivity across our search area) located within a circle of diameter 19$''$  (130 kpc), corresponding to a number density of $N(\smm >0.5\,{\rm mJy}) \approx 2 \times 10^4~\text{deg}^{-2}$.
By comparison, the average number of field sources with $S_{\rm 1.1\,mm} > 0.5$\,mJy within this area across all redshifts is less than one \cite{Ono2014}; thus, this field is overdense by more than a factor of 10.
When we account for the fact that all sources are at the same redshift, the volume density is  
$>1000\times$ the field density, assuming a redshift binning of $\Delta\rm z$=0.1 and the redshift distribution for SMGs\cite{Simpson2014a}. In Fig.~\ref{fig:comparison}, 
we plot `curves of growth' of the total 870-$\mu$m flux density versus on-sky area for \spt\ and other SMG-rich protoclusters  (see Methods for the details of the comparison sample). For \spt, we plot both the total flux density of the 14 confirmed protocluster members detected with ALMA and the total flux density of the extended LABOCA structure. The curve of growth for \spt\ rises much more steeply than those of the other high-redshift protoclusters, demonstrating its extreme density.
For \spt\, the on-sky area encompassing the accumulated 870-$\mu$m flux density (and thus approximately the total SFR) is as much as 3 orders of magnitude less than for other protoclusters at $z>2$. 
\spt\ clearly stands out as the densest collection of SMGs: although some other protoclusters contain as many SMGs, they extend over much larger areas on the sky, with separations often exceeding $10$ arcmin (22co-moving Mpc (cMpc) to 15 cMpc at $z = 4.3$ to $z = 2$). This comparison demonstrates that \spt\ is likely observed during a significantly more advanced stage of cluster formation than other high-redshift protoclusters, a  cluster core in the process of assembly rather than an extended structure that may not even collapse to form a cluster by the present day \cite{Chiang2013}. 

Also shown in Fig.~\ref{fig:comparison} is the maximal curve of growth predicted by a theoretical model for submm-luminous protocluster regions at $z\sim4.5$ (see Methods for details).
Except for \spt\ and the recent Herschel discovery SMM\,J004224\cite{Oteo2017}, the comparison high-$z$ protoclusters exhibit \ssubmm\ curves of growth fairly consistent with the model expectations.
The model prediction for the region spanned by \spt\ is $\sim10$\% of the observed total flux density of the 14 ALMA sources. The under-prediction is  more severe if we consider the extended LABOCA source: only $\sim5$\% of the observed flux density is recovered. This discrepancy may suggest that environmental effects (such as enhanced galaxy interactions or gas accretion in high-density environments) that are not included in the theoretical model employed are responsible for the extremely high SFR density exhibited by \spt. An alternative theoretical approach, `zoom' hydrodynamical simulations of protoclusters\cite{Granato2015}, can potentially capture such environmental effects, but to date, such simulations have been unable to reproduce the extremely high SFR inferred for \spt: of the 24 protocluster simulations presented by these authors, the maximum total SFR attained was $\sim 1700$ \mpy, an order of magnitude less than that of \spt.  However, the volume of the N-body simulation from which the 24 halos were selected was 1 $h^{-3}$ cGpc$^3$, which may be too small to contain an object as rare as \spt. 
Nevertheless, the existence of \spt, which contains an unprecedented concentration of rapidly star-forming SMGs when the Universe was only 1.4\,Gyr old,  poses a formidable challenge to theoretical models seeking to explain the origin and evolution of galaxy (proto)clusters.

 \spt\ may represent a significantly more advanced stage of cluster formation than the typical $z>4$ protoclusters identified to date, as outlined above.
 Since the cores of present-day galaxy clusters are characterized by massive elliptical galaxies with old-to-intermediate-age stellar populations \cite{Renzini2006} and SMGs are thought to be the high-redshift progenitors of present-day ellipticals \cite{Simpson2014a}, it is likely that at least some of the 14 SMGs located at the same redshift within a region $<130$ kpc in diameter will soon merge to form a massive elliptical galaxy at the core of a lower-redshift galaxy cluster. 

Theoretical studies have shown that at $z>4$, the progenitors of galaxy clusters should span $> 5$ comoving Mpc \cite{Chiang2013,Onorbe2013}, corresponding to an angular scale of as much as a degree; we are thus possibly observing only a small part of a much larger structure.
For \spt, it is unknown whether the overdensity extends over such a large scale, as more detailed observations are required to characterize the field surrounding \spt. We have demonstrated that the extended LABOCA-detected complex has submm colours similar to the core region identified by our ALMA observations and is thus likely all at $z\sim4.3$. We have also identified five additional bright SPIRE sources
in the surrounding  $\sim18 \times 18$\,cMpc field  
with similar red colours lying several arcmin from the core structure
 (see Methods). These are  candidates for additional protocluster members located in an extended, collapsing structure, similar to the comparison SMG overdensities shown in Fig.~\ref{fig:comparison}. If all these sources are confirmed to lie at $z = 4.31$, this would approximately double the far-IR luminosity of the cluster, making it by far the most active system known in the Universe.  
Since \spt\ was selected from a blind mm survey of 2500 deg$^2$ (approximately 1/16th of the sky),
it is unlikely there are more than approximately 16 such structures across the entire sky.
A full analysis of other unlensed sources from the SPT survey
to identify possible systems similar to \spt\ will place stronger constraints on early structure formation in the Universe.

\vskip 1cm
\noindent
{\bf Acknowledgments}\\
 This paper makes use of the following ALMA data: ADS/JAO.ALMA\#2016.0.00236.T
and ADS/JAO.ALMA\#2015.1.01543.T. ALMA is a partnership of ESO (representing its member states),
NSF (USA) and NINS (Japan), together with NRC (Canada) and NSC and ASIAA (Taiwan), in cooperation
with the Republic of Chile. The Joint ALMA Observatory is operated by ESO, AUI/NRAO and NAOJ.
This work is also based in part on observations
made with the Spitzer Space Telescope, which is operated by the Jet Propulsion Laboratory, California Institute
of Technology under a contract with NASA. The SPT is supported by the National Science Foundation
through grant PLR-1248097, with partial support through PHY-1125897, the Kavli Foundation and the Gordon
and Betty Moore Foundation grant GBMF 947.
This publication is based on data acquired with the Atacama Pathfinder
Experiment (APEX) under programme IDs  E-299.A-5045A-2017 and ID M-091.F-0031-2013. APEX
is a collaboration between the Max-Planck-Institut fur Radioastronomie, the European Southern Observatory, and the Onsala Space Observatory.
Supporting observations were obtained at the Gemini
Observatory, which is operated by the Association of Universities for Research in Astronomy, Inc., under
a cooperative agreement with the NSF on behalf of the Gemini partnership: the National Science Foundation
(United States), the National Research Council (Canada), CONICYT (Chile), Ministerio de Ciencia,
Tecnologa e Innovacion Productiva (Argentina), and Ministerio da Ciencia, Tecnologia e Inovacao (Brazil).
The Australia Telescope Compact Array (ATCA) is part of the Australia Telescope National Facility which is funded by the Australian Government for operation as a National Facility managed by CSIRO.
D.P.M., J.S.S., J.D.V., K.C.L., and S.J. acknowledge support from the U.S. National Science Foundation
under grant AST-1312950. 
S.C.C., T.B.M., and A.B. acknowledge support from NSERC. S.C.C. and T.B.M. acknowledge CFI and the Killam trust.
M.A. acknowledges partial support from FONDECYT through grant 114009.
The Flatiron Institute is supported by the Simons Foundation.
J.D.V. acknowledges support from an A.P. Sloan Foundation Fellowship.

\vskip 1cm
\noindent
{\bf Author Contributions} T.B.M. led the data analysis, and assembled the paper. S.C.C. designed the study, proposed the ALMA observations, reimaged the data, and analyzed the data products. C.C.H. developed the theoretical model and advised on the literature comparison. M.A. led the ATCA follow-up and the blind emission line studies.  A.W. procured and analyzed the deep LABOCA imaging.  M.B. provided the cluster mass and evolution context and discussion. J.S. reimaged the calibrated data. K.A.P. preform the SED fitting. T.B.M, S.C.C., M.A., K.A.P. and A.W. made the figures.  S.C.C., T.B.M., M.A., C.C.H., J.D.V., and A.W. wrote the manuscript. All authors discussed the results and provided comments on the paper. The authors are ordered alphabetically after A.W.

\vskip 1cm
\noindent
{\bf Author Information} The authors declare no competing financial interests. Correspondence and requests for materials should be addressed to T.B.M. at tim.miller@yale.edu

\clearpage
\setcounter{page}{1}
\textbf{\Huge Methods}
\section{Observations}

\subsection{SPT, LABOCA, and Herschel discovery and ALMA follow-up }

The South Pole Telescope\cite{Carlstrom2011} (SPT) possesses a unique combination of sensitivity, selection wavelengths (3, 2, and 1.4 mm), and beam size that potentially make it ideal for finding the active core regions of galaxy clusters forming at the earliest epochs.
Finding very distant ($z>4$), gravitationally lensed millimetre sources in the SPT survey is relatively straightforward, where the contrast to such distant bright sources  is high relative to the weak (generally undetected) galactic foregrounds
(Extended Data Figure~\ref{fig:spire_wide}).
However searching for the rare SMGs  in the SPT 2500 deg$^2$ survey that are unlensed, and therefore candidates for active groups and  protoclusters like \spt, involves sifting through the many gravitationally lensed sources, and typically involves multi-stage follow-up efforts using various facilities: a single dish mapping instrument like APEX-LABOCA to better localize the emission within the $\sim1'$ SPT beam, deep optical imaging to search for bright lensing galaxies, and high resolution ALMA mapping. 
The spatially extended sources in \spt\ found with LABOCA span more than an arcmin.
With deep upcoming surveys using the next generation SPT-3G receiver, this `extended-beam' thermal source structure may present a unique signature of many early forming protoclusters, affording the first complete census in the early epochs of structure formation.

A shallow, wide field SPIRE image over a 100 deg$^2$ subregion of SPT-SZ\cite{Ashby2013} reveals the red colours of
\spt, and that \spt\ appears to reside in something of a void in the $z\sim1$ foreground that dominates the SPIRE galaxy population.
However the high redshift of \spt\ means that it is not significantly brighter than many other SPIRE sources in this field, and aside from its colours, \spt\ does not stand out substantially from the field despite its extreme properties.
\spt\ is not detected in the all sky {\it Planck} survey\cite{Planck2014}, the lower sensitivity of {\it Planck} compared with SPT being exacerbated by beam dilution in the 3$'$ {\it Planck} beam.

Obtaining the redshift for \spt\ was beyond the scope of the original SPT-SMG redshift survey, due to the faintness of the unlensed components relative to the typical bright, gravitationally lensed SMGs found in the bulk of the SPT-SMG sample.
In ALMA Cy~0 \& 1, \spt\ was included in the 3-mm spectral scan redshift survey\cite{Vieira2013,Weiss2013},
but no lines were detected in the short $\sim1$ min integrations with $\sim$16 ALMA antennae. 
In Cy~3, a deeper follow-up 3-mm spectral scan was able to tentatively identify two CO lines and a double source structure with a likely redshift $z=4.30$, confirmed by APEX/FLASH C+ detection\cite{Strandet2016}.

\subsubsection{APEX - LABOCA}

We obtained 870-$\mu$m imaging of \spt\ using LABOCA on the APEX telescope. 
A shallow image with 1.6hr integration time was observed on 27 Sep 2010 
reaching $\sim5$\,mJy/beam rms. In August 2017 we obtained a deeper image 
(18.8h integration time, Project ID: E-299.A-5045A-2017, PI: Chapman) reaching a reaching a minimum noise level of 1.3\,mJy/beam and  $<2.0$ and $<1.5$\,mJy/beam rms for
75.3 and 32.4 sq arcmin, respectively (shown in Figure~\ref{fig:sources} \& Extended Data Figure~\ref{fig:laboca}). All 
observations were carried out using standard raster-spiral observations 
\cite{Siringo2009} under good weather conditions (PWV of 0.6\,mm and 0.8\,mm 
for the 2010 and 2017 observing campaigns, respectively). Calibration was 
achieved through observations of Uranus, Neptune and secondary calibrators and was found to 
be accurate within 8.5\% rms. The atmospheric attenuation was determined via skydips 
every 2hr as well as from independent data from the APEX radiometer which measures 
the line of sight water vapor column every minute. The data was reduced and imaged 
using the BoA reduction package \cite{Weiss2008}. 
LABOCA's central  frequency and beam size are 345\,GHz and 19.2$''$, resolving the SPT 1.4-mm elongated source into  two bright LABOCA sources.

Both LABOCA observations yield consistent calibration results with  peak intensity at 21$''$ 
resolution of 50\,mJy/beam for the brighter, southern component (RA 23:49:42.70, DEC -56:38:23.4).  
In addition the LABOCA map reveals a second source to the north at RA: 23:49:42.86, DEC: -56:37:31.02 with
a peak flux density at 21$''$ resolution of 17\,mJy/beam. Both sources are clearly extended even at
LABOCA's relatively coarse spatial resolution with observed source size of 28$''$$\times$25$''$ and 31$''$$\times$24$''$
for the sourthern and northern source, respectively. These components are connected by a faint bridge
emission. The total 870-$\mu$m flux density of the \spt\ system is
110.0$\pm$9.5\,mJy, of which $\sim$77\,mJy are associated with the southern component, $\sim$25\,mJy  with the northern
component, and $\sim$7\,mJy with the connection between the components (using the sub apertures shown in Extended Data Figure~\ref{fig:laboca}).
One additional submm source is detected at $>5\sigma$ in the LABOCA image to the east of the primary source, but having blue colours inconsistent with $z\sim4$, and not likely being a member of the extended protocluster.

\subsection{ALMA}

Observations using ALMA Band-3 targeted the CO(4-3) line in \spt\ centred in the lowest frequency of the spectral windows adopted (86-88\,GHz),   taken  under a Cycle 3 program 2015.1.01543.T (PI: K.\ Lacaille). Data was taken on June 24th, 2016 with a 47 min integration time. The array used 36 antennas with baselines ranging from 15 to 704 m, and provided a naturally weighted synthesized beam size of $\sim1''$. Pallas and J2343-5626 were used to calibrate the flux and phase respectively. Data was processed using the standard ALMA pipeline using natural beam weighting.

ALMA Band-7 imaging (276\,GHz)  were obtained under a Cycle 4 program (2016.0.00236.T; PI: S.\ Chapman) targeting the peak of the brightest LABOCA  source. Observations were obtained on December 14th, 2016 in a 40-2 array configuration with baseline lengths of 15-459 m, giving a naturally weighted synthesized beam size of $\sim1''$. There were 40 antennas available, with total on source integration time of 22 minutes. Ceres and J2357-5311 were used as flux and phase calibrators respectively. 
 The [CII] line ($\nu_{rest}=1900.5$\,GHz) was observed at as part of the same ALMA project on March 23rd, 2017, tuning in Band~7 to the redshifted line at $\nu_{ons}=358.3$\,GHz  in the upper sideband covering 356 to 360\,GHz. These observations used the 40-2 array configuration with baselines of 16-459 m, giving a naturally weighted synthesized beam size of $\sim0.5''$.  An on-source integration time of 14 min was obtained, and  J2357-5311 was used as both the flux and phase calibrator.
The data were re-processed using CASA  and the standard ALMA-supplied calibration using natural beam weighting to maximize sensitivity.

One dimensional spectra are extracted from the centroid of the line emission for each source and binned into 75 km s$^{-1}$ channels. Spectra are presented in Figure~\ref{fig:sources}, and are smoothed using a Gaussian filter with FWHM = 100 km s$^{-1}$ for presentation. A Gaussian line profile is fit using a least-squares method, providing errors to the velocity offsets from $z=4.304$ in Table~\ref{tab:phys} and line widths in Extended Data Table 2. The continuum level is left as a free parameter in the fitting function which is then subtracted to derive line fluxes and for presentation.

\subsubsection{Blind search for [CII]}
  
 We performed a blind search for [CII] line emission in the ALMA band 7 data cube toward \spt. For this, we follow the procedure used to detect line emitters in the ASPECS survey\cite{Aravena2016}. We use a data cube channelized at 100 km/s, without primary beam correction and continuum subtraction.
We used the Astronomical Image Processing System (AIPS) task {\sc SERCH}. This task convolves the data cube along the frequency axis with a Gaussian kernel defined by different input linewidths, subtracts surrounding continuum, and reports all channels and pixels that have a signal-to-noise ratio (SNR) over a specified limit. The SNR is defined as the maximum significance level achieved after convolving over the Gaussian kernels. We used a set of different Gaussian kernels, from 200 to 600 km/s and searched for all line peaks with SNR$>$4.0.

Once all peaks were identified, we used the IDL routine {\sc CLUMPFIND}\cite{Williams2011} to isolate individual candidates. A full list of 68 positive line peaks with SNR$>$4.0 were thus obtained. We quantified the reliability of our line search based on the number of negative peaks in our ALMA cube, using the same line procedure. We find 43 negative peaks with SNR$<$5.8 and none at a higher SNR. This means that all positive line candidates with SNR$>$6.0 are likely real (100\% purity). Out of the 14 [CII] line candidates detected, all have SNR$>$6.3 and 12 are associated with continuum detections in the ALMA data. 

\subsection{ATCA CO(2-1)}

\subsubsection{Observations}
We used the Australia Telescope Compact Array (ATCA) in its H168 hybrid array configuration to observe the CO(2-1) emission line ($\nu_{\rm rest} = 230.5380$\,GHz) toward \spt\ (with a primary beam size of 53$''$). The observations were performed as part of project ID C2818 during 2016 October 2,3 and 11 under good weather conditions (atmospheric seeing values 90-400 m) with five working antennas.

We used the ATCA 7-mm receivers, with the Compact Array Broadband Backend configured in the wide bandwidth mode \cite{Wilson2011}. This leads to a total bandwidth of 2\,GHz per correlator window and a spectral resolution of 1 MHz per channel (6.9 km/s per channel). The spectral windows were centred at observing frequencies of 43.5 and 45.0\,GHz, and aimed at observing the CO line and continuum emission, respectively. 

Gain and pointing calibration were performed every 10 min and 1 h, respectively.The bright sources 1921-293, 1934-638 and 2355-534 were used as bandpass, flux and gain calibrators, respectively. We expect the flux calibration to be accurate to within 15 per cent, based on the comparison of the Uranus and 1934-638 fluxes. The software package MIRIAD \cite{Sault1995} and the Common Astronomy Software Applications (CASA \cite{McMullin2007}) were used for editing, calibration and imaging.

The calibrated visibilities were inverted using the CASA task CLEAN using natural weighting. No cleaning was applied given the relatively low significance of the CO line detection in individual channels. The final data cube, averaged along the spectral axis, yields an rms of 0.23\,mJy beam$^{-1}$ per 100 km/s channel with a synthesized beam size of 5.6$''$ $\times$ 4.5 $''$ (PA=70.4 deg) at 43.5\,GHz.

\subsubsection{Results}

One source formally detected at the centre, which corresponds to CII/continuum sources B+C+G. This central CO source (C) is unresolved at the resolution of the ATCA observations. Other two sources are marginally detected to the West (w) and North (N) of the central source, coinciding with the location of CII/continuum sources 
D+E and A+K, respectively. We extracted spectra at these locations and obtained integrated line intensities, by fitting Gaussian profiles to the identified line emission. 

We compute CO luminosities using the integrated line intensities and compute gas masses by assuming a ULIRG X$_{CO}$ factor of 0.8 (M$_\odot$ (K km/s pc$^2$)$^{-1}$) and that the CO gas is in local thermodynamic equilibrium thus L$'_{({\rm CO}2-1)}\sim$L$'_{({\rm CO}1-0)}$~\cite{Solomon1997}. The results of the CO line observations are summarized in Extended Data Table~\ref{tab:atca}.
Collapsing the line-free spectral window along the spectral axis over the 2-GHz bandwidth, leads to a non-detection of the continuum emission down to 80\,$\mu$Jy/beam (3$\sigma$). 

These results confirm the finding from CO(4-3) line that that the main reservoir (72\%) of molecular gas resides in the B+C+G system, with  a smaller fraction hosted at the West and North locations. 

\subsection{Spitzer imaging}

This field was twice observed at 3.6 and 4.5\,$\mu$m with the Infrared Array Camera (IRAC\cite{Fazio2004}) on board the {\sl Spitzer Space Telescope}\cite{Werner2004}.  It was first observed in 2009 August as part of a large program to obtain follow-up imaging of a large sample of SPT-selected SMGs sources (PID 60194, PI Vieira).   The observing scheme used for PID\,60194 was to obtain 36 dithered 100\,sec integrations at 3.6\,$\mu$m and, separately, a much shallower $12\times30$\,sec integration at 4.5\,$\mu$m.  Later, in Cycle 8, the field was covered serendipitously as part of the {\sl Spitzer}-SPT Deep Field survey (PID 80032, PI Stanford; Ashby et al.\ 2013).  PID\,80032 surveyed 92\,deg$^2$ uniformly in both IRAC passbands to a depth of $4\times30$\,sec.  Using established techniques, we combined all exposures covering the SPT target from PID\,60194 and 80032 at 3.6 and 4.6\,$\mu$m to obtain the best possible S/N in our final mosaics, which were pixellated to 0.5$''$. Nine of the 14 sources identified by ALMA are detected in the IRAC bands at $>3\sigma$ in at least one of the 3.6 or 4.5\,$\mu$m channels, as shown in Extended Data Figure~\ref{fig:IRAC}.

\subsection{Analysis  of the surrounding field with SPIRE and  LABOCA  imaging }

	In Extended Data Figure~\ref{fig:spire}, our deep SPIRE RGB image is shown with LABOCA contours overlaid. A source sample is culled from the 250\,$\mu$m-selected catalog (135 sources with SNR(250\,$\mu$m)$>$3 in an area of 52 arcmin$^2$), where the source peaks are best defined. To account for the large beam size difference with SPIRE (ranging from 36$''$ at 500\,$\mu$m to 18$''$ at 250\,$\mu$m), 
we employed a deblending code,  using the 250\,$\mu$m positions as spatial priors, which provides the standard parameters as well as the covariance matrices highlighting the degeneracies (almost none at 250\,$\mu$m, but significant at 500\,$\mu$m). The code, FASTPHOT\cite{Bethermin2010}, takes into account these degeneracies to estimate the flux measurement errors.

	Colour-colour (CC) and colour-flux (CF) diagrams are shown in Extended Data Figure~\ref{fig:spire}. The CC diagram shows a 250\,$\mu$m-selected sample with SNR(250\,$\mu$m)$>$3 and is dominated by the $z\sim1$ cosmic infrared background (blue, green colours) in the {\it foreground} of \spt. 
	The CF diagram shows an additional SNR(500\,$\mu$m)$>$3 cut to highlight just the well detected 500\,$\mu$m source sub-sample.
		These diagrams highlight the extreme and red properties of \spt, but make clear that one of the three 250\,$\mu$m-peaks within the \spt\ LABOCA structure is very likely a foreground galaxy (green symbol highlighted in the figure shows very blue colours). Nevertheless, a full ALMA mapping of the structure is warranted given the uncertainties involved in the SPIRE deconvolution procedure. 

Five red sources consistent with $z\sim4$ ($S_{\rm 500\,\mu m} > S_{\rm 350\,\mu m} > S_{\rm 250\,\mu m}$) are found in the surrounding $\sim10$$'$$\times$10$'$ field and are  candidates for additional protocluster members in an extended, collapsing structure. If all these sources were bona fide $z=4.3$ sources, this would significantly increase the total 870-$\mu$m flux density (and thus the far-IR luminosity) of the cluster beyond the 110\,mJy found in the central structure, making it by far the most active system known in the Universe (see Figure~\ref{fig:comparison}).The deep LABOCA map marginally detects the closest of the five red SPIRE sources at $\sim 3 \sigma$, consistent with expectations given the SPIRE flux densities.  Full analysis of these surrounding SMGs will require additional follow-up efforts.

\section{Properties, Comparisons, Simulations}

\subsection{Derivation of physical properties}

We briefly describe our procedures for calculating various physical quantities from observables below. To derive SFR, we measure 870-$\mu$m flux density directly in the lower sideband (line-free  bands) of our ALMA Band-7 observations from Cycle~4, finding consistent measurements with those found in previous shallower observations\cite{Spilker2016}. We adopt an SFR-to-$S_{\rm 870\,\mu m}$ ratio of 150 $\pm$ 50 \mpy/mJy, which is typical for SMGs \cite{Barger2014}. The uncertainty in this ratio owes to variations in the dust temperature distribution amongst the SMG population, which are primarily driven by differences in the ratio of the luminosity absorbed by dust to the total dust mass\cite{Safarzadeh2015}. This combined with the measurement error dictates the error on the SFR shown in Table~\ref{tab:phys}

Gas mass is calculated from the CO(4-3) line luminosity, which is converted to CO(1-0) luminosity using a ratio between the brightness temperatures of these lines $r_{4,1}~=~0.41\pm 0.07$ found from the average of a sample of unlensed SMGs with multiple CO line transitions detected\cite{Bothwell2012}. We use a conservative conversion factor $\alpha_{CO} = 0.8 \frac{\text{\msun}}{\rm K\,km\,s^{-s}\,pc^2}$ and multiply by 1.36 to account for the addition of helium. 
When CO(4-3) is not significantly detected, we use our [CII] line luminosity and the average CO(4-3)/[CII] ratio for our detected sample; we denote these sources with asterisks in Table~\ref{tab:phys}.

\subsection{Spectral energy distribution of \spt}

The SPT, LABOCA and SPIRE measurements resolve the \spt\ structure to varying degrees, but none can isolate the core region resolved by our current ALMA observations with any confidence. We thus assemble a photometric catalog of the total \spt\ flux density from 250\,$\mu$m to 850\,$mu$\,mm and model the resulting total SED to estimate some global properties of the system. Although SPT does measure the flux at 1.4, 2.0 and 3.0 \,mm (27.9 $\pm$ 4.6 mJy, 5.2 $\pm$ 1.1 mJy and 0.5 $\pm$ 0.1 mJy, respectively), we do not include these points in the SED fit since the measurements are uncertain due to the elongated structure of \spt\ and difficulty with the filtering used to make the map. At IRAC wavelengths in the mid-infrared, we identify 9 SMGs detected at $>3\sigma$, and include the sum in the SED fit. We defer further analysis to followup work with forthcoming deeper {\it Spitzer}-IRAC data and follow up optical and NIR photometry.

We have used Code Investigating GALaxy Emission (CIGALE) \cite{Burgarella2005,Noll2009} for the SED fitting of the combined photometry of the source. The SED modelling assumes a single-component star formation history and solar metallicity \cite{Bruzual2003}. A Chabrier~\cite{Chabrier2003} IMF is assumed. The resulting best-fitting SED is shown in Extended Data Figure~\ref{fig:SED}. The IR luminosity (from 8--1100\,$\mu$m) is $(8.0\pm1.0)\times10^{13}$ L$_\odot$.

\subsection{Protocluster comparison sample}
To place \spt\ in context and compare to other systems claimed to be protoclusters, we assemble from the literature various SMG-rich overdensities  at $2<z<5$. 
Although a direct comparison of the number counts (number/deg$^2$) of SMG-overdense systems can be performed, it involves making somewhat arbitrary choices of enclosed areas and redshift boundaries.  We have opted in Figure~\ref{fig:comparison} to instead show a curve of growth analysis of the 870-$\mu$m flux density.
 Only galaxies confirmed to be protocluster members via spectroscopic redshifts are considered.  
The data are drawn from a recent compilation\cite{Casey2016} and original references therein.

The GOODS-N overdensity  at $z=1.99$ \cite{Chapman2005,Chapman2009,Blain2004} spans a  
$\sim$ 10$^{\prime}$ by 10$^{\prime}$ field in the Hubble Deep Field North containing 9 SMGs in $\Delta$z = 0.008. The probably of finding this large of an overdensity being drawn from the field distribution by chance is $<0.01\%$. Interestingly, only a modest overdensity of Lyman-break galaxies is found in this GOODS-N structure. 

The COSMOS $z=2.5$ SMG overdensity \cite{Casey2015} is similar to the GOODS-N structure in terms of the numbers and luminosities of the component SMGs, the angular size of the system, and the modest overdensity of LBGs associated with it.

The MRC1138 overdensity was originally discovered as an overdensity of Ly$\alpha$ and H$\alpha$ 
emitters\cite{Kurk2000}. Follow-up observations\cite{Kuiper2011,Dannerbauer2014} revealed the presence of 5 SMGs, in addition to an AGN known as the `Spiderweb galaxy'. This is a radio-loud AGN that resides in a large Ly-$\alpha$ halo. 

The SSA22 protocluster was one of the first discovered by  observing an overdensity of LBGs\cite{Steidel1997}. It is an extremely extended structure located at $z=3.09$, with LAEs spanning greater than 50 comoving Mpc (cMpc)\cite{Hayashino2004}. Submm observations of the field have revealed a population of at least eight SMGs \cite{Chapman2001,Chapman2005,Geach2005,Tamura2009,Umehata2015} .

The COSMOS $z=2.1$ protocluster\cite{Hung2016} lacks sufficiently deep 850-$\mu$m data to characterize the {\it Herschel}-SPIRE sources identified in the structure. We estimate 870-$\mu$m flux densities by taking their published $L_{\rm IR}$ (integrated over 3--1100\,$\mu$m) and use the SED of Arp 220  to estimate \ssubmm, finding that $L_{\rm IR} = 2 \times 10^{12}$ \lsun corresponds to $S_{870\,\mu m}$ = 1\,mJy at $z\sim2$. For the SSA22 protocluster, we use the measured 870-$\mu$m flux density when available and otherwise estimate it from the 1.1-mm flux using a standard conversion at $z\sim3$ of  S$_{870\,\mu m}$ = 2$\times$ S$_{\rm 1.1\,mm}$. To create the curves of growth for 
Figure~\ref{fig:comparison}, the centre of each protocluster is defined by computing the median RA and DEC of all submm sources. We checked that adjusting the centres of the curve of growth tracks randomly by $\sim1'$ did not boost the curves  by more than 10\%, demonstrating that the curves of growth for the literature SMG overdensities are insensitive to the adopted centre.

Recently, there have also been detections of SMG overdensities at $z>4$. The first, GN20, at $z = 4.05$, was discovered through the serendipitous detection of CO(4-3) from two SMGs\cite{Daddi2008}, with two further SMGs detected subsequently\cite{Daddi2009}.  An excess of $B$-band dropouts is also observed in this structure, several of which are spectroscopically confirmed to lie at $z\sim4.05$.
HDF850.1 contains  a single SMG, a QSO, and 11 spectroscopically confirmed galaxies. The SMG has a confirmed redshift of $z=5.18$\cite{Walter2012}. The \textit{AzTEC-3} overdensity is centred on a single SMG at $z=5.3$, with 12 spectroscopically confirmed optical galaxies at the same redshift. This is a relatively dense structure, with most of the galaxies residing within a circle $\sim$1$^{\prime}$ in diameter. 
The most luminous example at $z>4$, SMM\,J004224, was recently found from the {\it Herschel} surveys\cite{Oteo2017}, with several additional 870$\mu$m sources in the surrounding field. In Fig. 2 we plot, as an extension, the flux from the satellites which have observed SPIRE colors consistent with $z=4$\cite{Lewis2017}.

Overdensities of SMGs and optical galaxies have also been found around high-redshift radio galaxies (HzRGs)\cite{Noirot2016}, continuing to confirm HzRGs as useful beacons of structure forming in the early Universe. However none of these systems come close to the level of overdensity found in \spt, and furthermore, they suffer from the bias inherent in targeting these sources, namely, that one or more protocluster members have to be radio-luminous.

There have also been discoveries of compact binary HyLIRG systems, the most luminous of which  is the $z=2.4$ source HATLAS J084933 \cite{Ivison2013}, with others approaching this luminosity\cite{Fu2013,Chapman2015}. In each of these systems, the dynamics and SFRs are dominated by two SMGs, but there is no strong evidence of any surrounding protocluster in the form of an excess of galaxies selected optically or in the submm. In one case\cite{Chapman2015}, there is evidence for a relative void around the structure. These systems may simply be instances of very rare events in fairly typical (but still massive) halos\cite{Miller2015}, analogous to hyper-luminous quasars\cite{Trainor2012}.

Theoretical studies of N-body simulations have shown that the progenitors of $z=0$ dark matter halos with masses $>10^{15}$ \msun\ should extend to effective radii of $\gtrsim 5$ cMpc at $z>2$.\cite{Chiang2013,Onorbe2013} Since the overdensities listed above are mostly concentrated in small areas, it is difficult to asses their exact evolution or compare them easily to simulated structures.  Interpreting a small overdense region at high-redshift as a `protocluster core' is certainly prone to misinterpretation, and small overdense regions at high redshift can evolve into halos spanning a range of masses at the current epoch\cite{Chiang2013}. These authors suggest investigating if an overdensity extends to larger scales ($> 20$ cMpc) to better determine whether it will form an $M \gtrsim 10^{15}$\msun cluster. However this is difficult at high redshift because the excess of galaxies will be less pronounced on larger scales, and it is challenging to  detect high-redshift, low-luminosity galaxies.

To assess the evolution of \spt\ we also compare to a selection of the highest mass galaxy clusters at lower redshift (z $\lesssim 2.5$). XMMU J2235.3\cite{Rosati2009} and XLSSC 122\cite{Mantz2017} were both discovered in x-ray surveys with masses derived from the x-ray light profile.IDCS J1426.5+3508\cite{Stanford2012,Brodwin2016}, JKCS5041\cite{Andreon2014}, CL J1449+0856\cite{Gobat2011,Gobat2013} and CL J1001+0220\cite{Wang2016} were discovered as overdensities of massive red galaxies and have corresponding x-ray detections. Masses for all are estimated using these x-ray detections and consistent with dynamical masses estimated using the velocity dispersion. We also show the sample of clusters discovered in the 2500 square degree SPT survey using the Sunyaev-Zel'dovich effect, with mass estimate based on the velocity dispersions\cite{Bocquet2014}.

\subsection{Geometry and Dynamics}

In Extended Data Figure~\ref{fig:dv_dr} we investigate the geometry and kinematics of the \spt\ system. By analysing $\Delta \rm V$ vs. projected radius for our sample of 14 galaxies it is clear that, even with our most conservative mass estimate, at least 12 of the 14 galaxies appear bound. We show the escape velocity as a function of radius for a point source and NFW profile with a mass of $1.16 \times 10^{13} \rm M_\odot$. The NFW profile assumes a virial radius of 200 kpc and concentration of 5, typical values of massive halos at this epoch, found in N-body simulations described in \S~2.5. Even if the projected radius only accounts for a fraction of the true physical distance (i.e. separated along the line of sight) it is likely that most of the galaxies with low relative velocity ($\Delta \rm V < 500\ km\,s^{-1}$) will still be bound and eventually collapse into a single galaxy. The cumulative distribution of relative velocities is also shown, alongside that of a Gaussian distribution with $\sigma = 408 \rm km\,s^{-1}$.The observed distribution is smooth and well fit by a Gaussian.  The relative kinematics of the 14 galaxies is consistent with most, if not all, members being mutually bound with \spt\ following a Gaussian $\Delta \rm V$ distribution characteristic of a virialized system.

An alternative interpretation of this system is that we are observing individual galaxies, or  separate less massive halos  along a filament aligned with our line of sight. The probability of this occurring is potentially boosted by our selection technique, since the beam size of SPT is approximately 1$'$. We are preferentially sensitive to structure on this scale, and  require several of the most luminous galaxies found in the Universe together in the SPT beam to exceed our detection threshold. 

To explore this possibility,  we also show in Extended Data Figure~\ref{fig:dv_dr} the most extreme interpretation of \spt\ where none of the velocity offsets are peculiar, and the physical distribution of the \spt\ galaxies is represented by the relative velocities being entirely due to cosmic expansion rather than peculiar motions. In this case the galaxies of course are maximally separated along a (proper distance) 3Mpc {\it filament} compared to their 130kpc maximum tangential extent. However, this cannot be the true distribution, and a more realistic interpretation would entail some $\sim$20-30\% of velocities being Hubble flow (600-900pc extent), and the remainder representing peculiar motions.  (Note that this would be true for any massive halo at this epoch given the expected velocity dispersion of $\sim 400\rm km\, s^{-1}$ and Hubble constant at this epoch of $470\rm km\, s^{-1} Mpc^{-1}$.) This explanation, if substantiated, would alter our interpretation of the system, specifically decreasing the mass estimate as the velocity offsets are not due to peculiar motions and the system is not virialized.

However, given the observed kinematics and spatial configuration of the galaxies discussed above, we argue that \spt\ is less likely to consist of multiple groups, widely separated on a line of sight filament. To further back up this claim, we perform a systemic search for similar filamentary structures in the N-body simulations described in \S~2.5. We first search for groups at $z>4$ with more than 8 galaxies  that have a projected radius of $< 150$ kpc but can be extended up to $5$ Mpc along the line of sight. There are three such systems  found in our 1 cGpc$^3$ simulation, and they are also displayed in E.D. Fig.~\ref{fig:dv_dr} showing their geometry. All three of these analogous systems in the simulation have total halo masses of $\sim 10^{13}\rm M_\odot$.
None are extended significantly past 500 kpc, implying that at $z\sim4$, there are no substantially extended \textit{filaments} hosting massive galaxies comparable to even the gas masses our \spt\ ALMA galaxies in the simulation.
These simulated structures that we have found are not filaments -- they are close to collapsed structures which are slightly cigar-shaped, and while \spt\ could in principle be distributed like this, it does not fundamentally change our discussion in the paper.
(We also note that  we are plotting 2 different things in E.D. Fig.~\ref{fig:dv_dr} 
velocity offsets for 2349 and actual geometry for the simulation galaxies.) 
These results further suggest that most of the velocity offsets in \spt\ galaxies are truly mostly due to peculiar motion rather than line of sight projection.
At least one of the simulation systems found appears to be chance projections of multiple groups, but it has a characteristically different $\Delta \rm V$ distribution. The system is found to have multiple smaller groups  with $\sigma \sim 200\, \rm km\,s^{-1}$, reflecting their smaller masses, separated in velocity space by $500\, \rm km\,s^{-1}$. 

We also note that a filamentary structure feeding a $\sim 10^{13}\rm\, M_\odot$ halo should have a characteristic width of 200 kpc - 400 kpc \cite{Mandelker2017,Cautun2014} compared to the 80 kpc (FWHM) width we find for our 14 ALMA galaxies. 
This is suggestive that even with a direct line-of-sight view down a filament, the configuration is far more concentrated than one would infer based on the typical size of filament.
We then ask what sort of environment a $\sim80$kpc wide filament (consistent with our \spt\ galaxies) would connect, and infer a typical halo mass of $\sim 2 \times 10^{12} \rm M_\odot$ halo \cite{Mandelker2017} .
Given that the \spt\ galaxies contain at least this much mass entirely in their cold gas (as we have adopted the  lowest plausible $\alpha_{CO}=0.8$ conversion factor), and are likely hosted by haloes at least twice as large as this, it becomes somewhat contradictory that they could be found along a filament that is connecting a significantly less massive halo.

A further important piece of evidence against these 14 galaxies being a filament is that we find no additional sources in the surrounding spectral windows in either side band, both in the band 7 and band 3 ALMA data (Extended Data Figure~\ref{fig:dv_dr} - d).
While all 14 sources are clustered within 1500 km s$^{-1}$ of each other, our full frequency coverage extends over   relative velocities of 6500 km s$^{-1}$ (for [CII]) and 27,000 km s$^{-1}$  (for CO(4-3)),  or many tens of Mpc line of sight distance. If \spt\ was actually an elongated filament, one might expect to see a  distribution of sources over a much larger fraction of our observed frequency bandwidth. Although we can not rule out the possibility that we are observing multiple galaxies along our line of sight, the observed geometry and kinematics suggest \spt\ consists of a single group of mutually gravitationally bound galaxies.

Another way to assess the configuration of this system is by employing the 
Millennium Simulation \cite{Springel2005a} database, using the
implementation of the \cite{Bower2006} galaxy-formation
recipe.  We search  the model output at $z\sim4$ for galaxies with a total baryonic mass in excess
of $\sim 10^{11}$\,M$_{\odot}$, consistent with the combined mass
of gas and stars estimated for the brightest ALMA-resolved galaxies in \spt.  We use
this total baryonic mass cut as it means we are less sensitive to the details of the
early star-formation histories of galaxies in the model.
We find only one galaxy in the $3.2\times 10^8$\,comoving-Mpc$^3$ simulation volume 
at $z=4$ with a
total baryonic mass above $10^{11}$\,M$_{\odot}$. Half of the
baryonic mass  in this model galaxy is in stars and it has a $3\times 10^8$\,M$_{\odot}$ black
hole. Moreover this galaxy is the central galaxy of a $1\times 10^{13}$\,M$_{\odot}$
halo,  the optimal environment to find merging galaxies
according to simulations\cite{Hopkins2008}. Searching the
environment of this system we find another four massive
galaxies are distributed across a $\sim$\,0.5-comoving-Mpc-diameter region
around the central galaxy, with baryonic masses of  $>0.7\times 10^{10}$\,M$_{\odot}$ 
($> 5$\% of the primary). % and predicted $K_{\rm AB}$ magnitudes of 22.2--22.7.
By redshift zero these galaxies are all predicted to reside in a  $1\times 10^{15}$\,M$_{\odot}$ halo, consistent with our other assessments of the outcome of this system.

\subsection{Simulations}
To further place \spt\ in context, we compare with the predictions of a theoretical model for SMG overdensities.\cite{Hayward2013,Miller2015}
We employ the MultiDark\cite{Prada2012} N-body simulation, which is one of the largest (2.91 Gpc$^3$) available N-body simulations 
that still resolves SMG-like halos ($M_{\rm halo} \gtrsim 10^{12} $ \msun).
The $z = 4.68$ and $z=4.1$ snapshots, which are the available snapshots closest in redshift to \spt, are analyzed. Halo catalogs were created using the Rockstar halo finder\cite{Behroozi2011}, and stellar masses are assigned to dark matter halos using a relation
derived based on sub-halo abundance matching relation\cite{Behroozi2013}.
To assign SFRs, it is assumed that the distribution of specific SFR (SFR per unit stellar mass, hereafter SSFR) is the sum of two Gaussians, corresponding to quiescently star-forming and starburst galaxies \cite{Sargent2012}.
The median SSFR value is based on the abundance-matching-derived relation\cite{Behroozi2013}, and the starburst fraction and the widths of the Gaussian distributions are set based on observations of massive, high-redshift star-forming galaxies similar to the members of \spt\cite{Sargent2012}. $M_{\text{d}}$ is estimated from stellar mass using empirical gas fraction and metallicity relations\cite{Hayward2013b}.
Once SFR, $\mstar$, and $M_{\text{dust}}$ values are assigned to each halo, \ssubmm\ is calculated using the following fitting function, which was derived based on the results
of performing dust radiative transfer on hydrodynamical simulations of both isolated and interacting galaxies\cite{Hayward2011,Hayward2013}:
\begin{equation}
S_{870\,\mu \text{m}} = 0.81 \text{ mJy } \left( \frac{\rm SFR}{100 \text{ \mpy}} \right)^{0.43}
\left(\frac{M_\text{d}}{10^{8} \text{ \msun}} \right)^{0.54},
\end{equation}
where \ssubmm\ is the 870-$\mu$m flux density, SFR is the star formation rate, and $M_{\text{d}}$ is the dust mass. 
Scatter of 0.13 dex is included when applying the relation. 

\begin{figure*}
	\centering
	\includegraphics[width=0.51\textwidth]{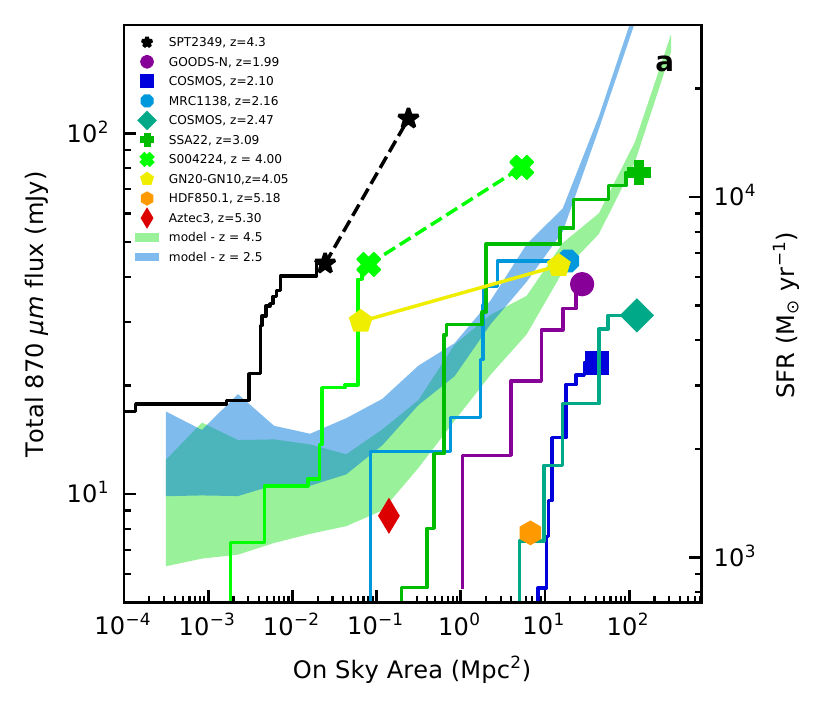}
	\includegraphics[width=0.48\textwidth]{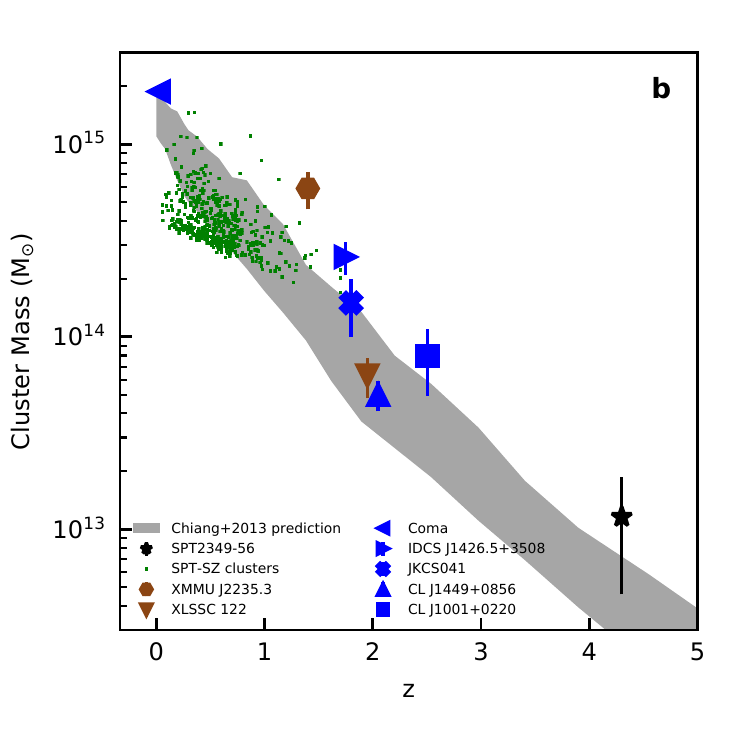}
	\caption{\textbf{Comparison of \spt\ to other cluster and proto-cluster systems.} 
	(a) The cumulative 870-$\mu$m flux density vs. on-sky area for \spt, compared to other SMG-rich overdensities at high redshift (see Methods for details). The solid black line  shows the ALMA-identified sources in \spt, while the dashed line includes the wider-field LABOCA-detected structure.
	The blue (green) shaded region denotes the maximum flux density vs. area curves obtained in 100 Monte Carlo realizations of a theoretical model for submm-luminous protoclusters at $z = 4.5$ ($z = 2.5$) based on an N-body simulation (see Methods for full details). Most of the literature SMG overdensities are consistent with the model expectations, whereas \spt\ lies vastly above the region spanned by the model. A recently discovered $z = 4$ protocluster from {\it Herschel}\cite{Oteo2017}, SMM\,J004224, is quite a unique system but $>10$ times less dense than and likely only $\sim 50$\% the total luminosity of \spt.
(b) The cluster mass versus redshift is shown for \spt\ and other massive galaxy clusters from the literature with detected ICM and well-defined masses. The colour scheme highlights the different methods for selecting massive clusters employed (brown = X-ray, blue = optical, green = Sunyaev-Zeldovich effect). Error bars represent the 1 $\sigma$ standard deviation. We also show the mean protocluster most-massive-progenitor mass vs. $z$ relation predicted by N-body simulations\cite{Chiang2013}. The location of \spt\ in this plane suggests a very massive descendant (halo mass of $>10^{15}$ \msun\ at $z$ = 0), although we caution that the complex growth histories of dark matter halos make it difficult to reliably predict the $z$ = 0 halo mass from the halo mass at a given epoch\cite{Cole2008}. 
	 }
	\label{fig:comparison}
\end{figure*}

Once \ssubmm\ has been assigned to each halo, we search the entire simulation volume for the most luminous regions. We begin at each independent halo and calculate the total \ssubmm\ of all halos within a given projected radius $r$ and LOS distance along a given axis of this halo. We use a line of sight distance of 1 Mpc for the $z=4$ snapshot and 2 Mpc for the $z=2.5$ snapshot reflecting the Hubble constant at each epoch and the expected velocity dispersion of $\sim 400\ \rm km\ s^{-1}$. For each value of $r$, we record the largest total \ssubmm\ obtained (across all halos). One hundred Monte Carlo iterations are performed for each snapshot; in each iteration, galaxy properties are re-assigned, drawing from the distributions described above. The shaded region in Figure~\ref{fig:comparison} shows the entire region spanned by the 100 realizations of the maximum \ssubmm\ vs. area curves. To compare to \spt\ to lower redshift proto-clusters we preform a similar analysis on a snapshot at $z = 2.49$ with 20 Monte Carlo iterations.

%\bibliographystyle{naturemag}
%\bibliography{library}
	
\begin{table}[!h]
\caption{Derived physical properties of SPT2349-56 protocluster members.}
\label{tab:phys}
\begin{center}
\begin{tabular}{c|ccc}
Source & $\Delta V^{\ddagger}$ & SFR  & $M_{\rm gas}$\\
 & [km \,s$^{-1}$] & [\msun \,yr$^{-1}$] & [$10^{10}$ \msun]  \\ \hline
A & $-$90 $\pm$ 35 & 1170 $\pm$ 390 & 12.0 $\pm$ 2.1  \\
B & $-$124 $\pm$ 31 & 1227 $\pm$ 409 & 11.2 $\pm$ 2.0  \\
C & 603 $\pm$ 12 & 907 $\pm$ 302 & 6.7 $\pm$ 1.2 \\
D & $-$33 $\pm$ 40 & 530 $\pm$ 182 & 8.4 $\pm$ 1.5  \\
E & 84 $\pm$ 21 & 497 $\pm$ 179 & 4.8 $\pm$ 0.9  \\
F & 395 $\pm$ 82 & 505 $\pm$ 169 & 3.4 $\pm$ 0.7  \\
G & 308 $\pm$ 42 & 409 $\pm$ 137 & 2.9 $\pm$ 1.3$^{\dag}$  \\
H & $-$719 $\pm$ 28 & 310 $\pm$ 105 & 4.4 $\pm$ 2.0$^{\dag}$  \\
I & 310 $\pm$ 78 & 268 $\pm$ 91 & 2.2 $\pm$ 0.5  \\
J & $-$481 $\pm$ 35 & 243 $\pm$ 85 & 2.2 $\pm$ 0.5  \\
K & 631 $\pm$ 12 & 208 $\pm$ 71 & 3.1 $\pm$ 1.4$^{\dag}$  \\
L & $-$379 $\pm$ 18 & 122 $\pm$ 43 & 3.3 $\pm$ 1.5$^{\dag}$  \\
M & 34 $\pm$ 21 & 75 $\pm$ 34 & 1.2 $\pm$ 0.6$^{\dag}$ \\
N & 90 $\pm$ 25 & 64 $\pm$ 29 & 1.0 $\pm$ 0.5$^{\dag}$  \\ \hline
\end{tabular}
\end{center}
\vspace{1ex}
{\scriptsize
$^{\ddag}$ Velocity offsets relative to the mean redshift, z = 4.304\\
$^{\dag}$ [CII] line used to derive \mgas as CO(4-3) is not detected\\
}
\end{table}

\clearpage
\setcounter{table}{0}
\setcounter{figure}{0}
\renewcommand{\figurename}{\textbf{Extended Data Figure}}
\renewcommand{\tablename}{\textbf{Extended Data Table}}

\begin{table}
\caption{Observed properties of SPT2349-56 protocluster members}
\label{tab:obs}
\resizebox{\columnwidth}{!}{
\begin{tabular}{c|cccccccc}
\centering
Source & RA (J2000)	& Dec (J2000) & $S_{\rm 1090\,\mu m}$ & $S_{\rm 870\,\mu m}$ & CO(4-3) $\int$ S dv& CO(4-3) $\sigma_{\rm V}$& [CII] $\int$ S dv & [CII] $\sigma_{\rm V}$ \\
{} & [h:m:s] & [d:m:s]	& [mJy] & [mJy] & [Jy km s$^{-1}$] & [km s$^{-1}$]  & [Jy km s$^{-1}$] & [km s$^{-1}$]   \\ \hline
A & 23:49:42.67 & $-$56:38:19.3 & 4.63 $\pm$ 0.04 & 7.8 $\pm$ 0.1 & 0.99 $\pm$ 0.03 & 376 $\pm$ 46 & 8.81 $\pm$ 0.26 & 354 $\pm$ 30  \\
B & 23:49:42.79 & $-$56:38:24.0 & 4.35 $\pm$ 0.04 & 8.2 $\pm$ 0.1 & 0.92 $\pm$ 0.03 & 341 $\pm$ 38 & 7.53 $\pm$ 0.22 & 314 $\pm$ 28 \\
C & 23:49:42.84 & $-$56:38:25.1 & 2.69 $\pm$ 0.04 & 6.0 $\pm$ 0.1 &  0.55 $\pm$ 0.02 & 154 $\pm$ 13 & 4.43 $\pm$ 0.17 & 160 $\pm$ 10 \\
D & 23:49:41.42 & $-$56:38:22.6 & 2.20 $\pm$ 0.08 & 3.5 $\pm$ 0.3 & 0.69 $\pm$ 0.04 & 485 $\pm$ 64 & 3.62 $\pm$ 0.78 & 346 $\pm$ 129  \\
E & 23:49:41.23 & $-$56:38:24.4 & 2.12 $\pm$ 0.11 & 3.3 $\pm$ 0.4 & 0.39 $\pm$ 0.02 & 199 $\pm$ 23 & 3.47 $\pm$ 1.24 & 310 $\pm$ 137  \\
F & 23:49:42.14 & $-$56:38:25.8 & 1.69 $\pm$ 0.05 & 3.4 $\pm$ 0.1 &  0.28 $\pm$ 0.03 & 396 $\pm$ 103 & 4.28 $\pm$ 0.35 & 353 $\pm$ 35 \\
G & 23:49:42.74 & $-$56:38:25.1 & 1.11 $\pm$ 0.04 & 2.7 $\pm$ 0.1 & - & - & 2.45 $\pm$ 0.23 & 305 $\pm$ 50 \\
H & 23:49:43.46 & $-$56:38:26.2 & 0.85 $\pm$ 0.05 & 2.1 $\pm$ 0.1 & - & - & 3.63 $\pm$ 0.30 & 236 $\pm$ 31  \\
I & 23:49:42.22 & $-$56:38:28.3 & 0.78 $\pm$ 0.05 & 1.8 $\pm$ 0.1 & 0.18 $\pm$ 0.03 & 277 $\pm$ 90 & 3.18 $\pm$ 0.32 & 236 $\pm$ 24  \\
J & 23:49:43.22 & $-$56:38:30.1 & 0.61 $\pm$ 0.06 & 1.6 $\pm$ 0.2 & 0.19 $\pm$ 0.02 & 151 $\pm$ 38 & 3.79 $\pm$ 0.29 & 138 $\pm$ 15  \\
K & 23:49:42.96 & $-$56:38:17.9 & 0.34 $\pm$ 0.04 & 1.4 $\pm$ 0.1 &  - & - & 2.54 $\pm$ 0.17 & 129 $\pm$ 12  \\
L & 23:49:42.38 & $-$56:38:25.8 & 0.23 $\pm$ 0.04 & 0.8 $\pm$ 0.1 & - & - & 2.78 $\pm$ 0.20 & 176 $\pm$ 20 \\
M & 23:49:43.39 & $-$56:38:21.1 & 0.21 $\pm$ 0.05 & 0.5 $\pm$ 0.2 & - & - & 1.04 $\pm$ 0.14 & 87 $\pm$ 23 \\
N & 23:49:43.27 & $-$56:38:22.9 & 0.18 $\pm$ 0.04 & 0.4 $\pm$ 0.1 & - & - & 0.86 $\pm$ 0.16 & 128 $\pm$ 26  \\ \hline
\end{tabular}
}
\end{table}

\begin{table}
	\centering
	\caption{Properties of the 3 ATCA CO(2-1) sources}
	\label{tab:atca}
	\begin{tabular}{c|ccccc}
		
		ATCA source & ALMA ID & $\int$ S dv  & $\sigma_V$  & L$^{\prime}$(CO 2-1)  & $M_{\rm gas}$\\
		{} & {} & [Jy km s$^{-1}$] & [km s$^{-1}$] & $10^{11}$ [K km s$^{-1}$ pc$^{2}$] & [$10^{11}$ \msun] \\ \hline
		Central (C) & B, C, G & 0.69 $\pm$ 0.076 & 372 $\pm$ 47 & 1.22 $\pm$ 0.12 & 1.33 $\pm$ 0.15 \\
		West (W) & D, E & 0.16 $\pm$ 0.04 & 166 $\pm$ 47 & 0.29$\pm$ 0.07 & 0.32 $\pm$ 0.08 \\
		North (N)& A, K & 0.085 $\pm$ 0.0028 & 175 $\pm$ 68 & 0.15 $\pm$ 0.05 & 0.16 $\pm$ 0.05 \\
	\end{tabular}
\end{table}

\begin{table}
\caption{Observed properties of all red $(S_{\rm 500\,\mu m} > S_{\rm 350\,\mu m} > S_{250\,\mu m})$ SPIRE sources in the field surrounding SPT2349-56. The LABOCA sources corresponding to \spt\ are listed first, and the red SPIRE sources in the surrounding field follow. All sources listed are highlighted in Extended Data Figure~\ref{fig:spire}.}
\label{tab:spire}
\begin{center}
\begin{tabular}{ccccccc}
RA (J2000)	&	Dec (J2000)	& $S_{\rm 250\,\mu m}$ 	& $S_{350\,\mu m}$	& $S_{500\,\mu m}$ 	& $S_{850\,\mu m}$ 	& d$^{\dag}$ \\
{[h:m:s]}	&	{[d:m:s]} 	& [mJy] 				& [mJy] 			& [mJy] 			& [mJy] 			& [arcmin]\\
\hline
23:49:42 & $-$56:38:25 & 45 $\pm$ 3 & 71 $\pm$ 3 & 96 $\pm$ 3 & 77.0 $\pm$ 2.9 & - \\
23:49:43 & $-$56:37:31 & 21 $\pm$ 3 & 37 $\pm$ 3 & 43 $\pm$ 3 & 25.0 $\pm$ 2.8 & 0.9 \\
\hline
23:49:25 & $-$56:35:27 & 23 $\pm$ 4 & 26 $\pm$ 4 & 32 $\pm$ 4 & 2.9  $\pm$ 1.7 & 5.2 \\
23:49:39 & $-$56:36:33 & 12 $\pm$ 3 & 16 $\pm$ 3 & 23 $\pm$ 4 &  3.9 $\pm$ 1.3 & 2.1 \\
23:49:36 & $-$56:41:17 & 7 $\pm$ 3 & 14 $\pm$ 3 & 19 $\pm$ 3 &  3.2$\pm$ 1.6 & 3.2\\
23:49:55 & $-$56:34:17 & 6 $\pm$ 4 & 10 $\pm$ 3 & 20 $\pm$ 5 & 4.8 $\pm$ 1.8 & 5.3\\ 
23:49:12 & $-$56:40:31 & 11 $\pm$ 5 & 16 $\pm$ 5 & 22 $\pm$ 5 & 6.8 $\pm$ 2.6 & 7.7 \\
\hline
\end{tabular}
\end{center}
%\vspace{1ex}
{\scriptsize 
$^{\dag}$ Distance from central \spt\ source.
}
\end{table}

\begin{figure}
	\centering
	\includegraphics[width = 0.65\textwidth]{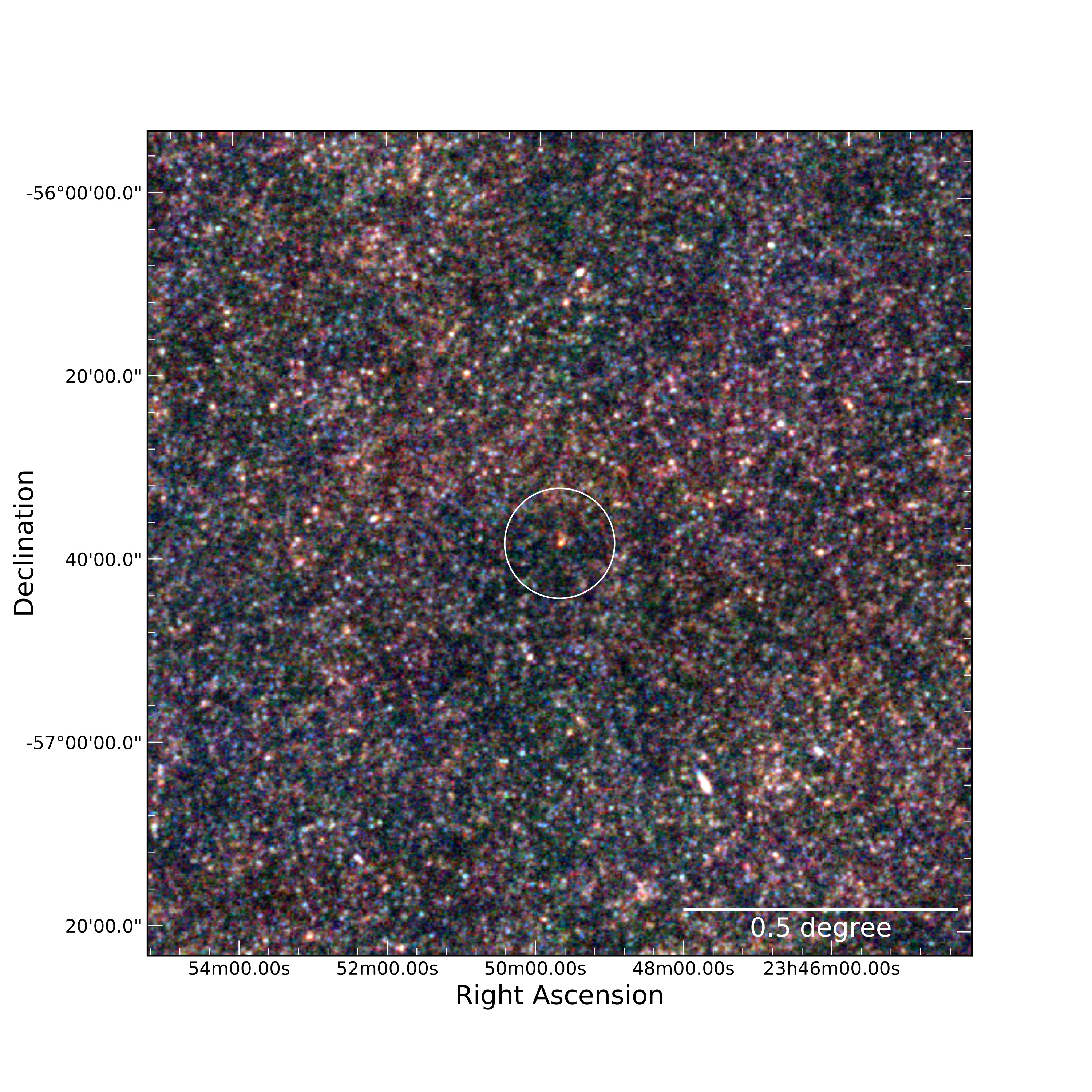}
	\caption{\textbf{{\it Herschel}-SPIRE image.} A RGB scale is used to represent 500, 350, and 250\,$\mu$m, with the red \spt\ extended complex clearly visible in a relative void in the foreground $z\sim1$ cosmic infrared background (blue to green-coloured galaxies).}
	\label{fig:spire_wide}
\end{figure}

\begin{figure}
	\centering
	\includegraphics[width = 0.95\textwidth]{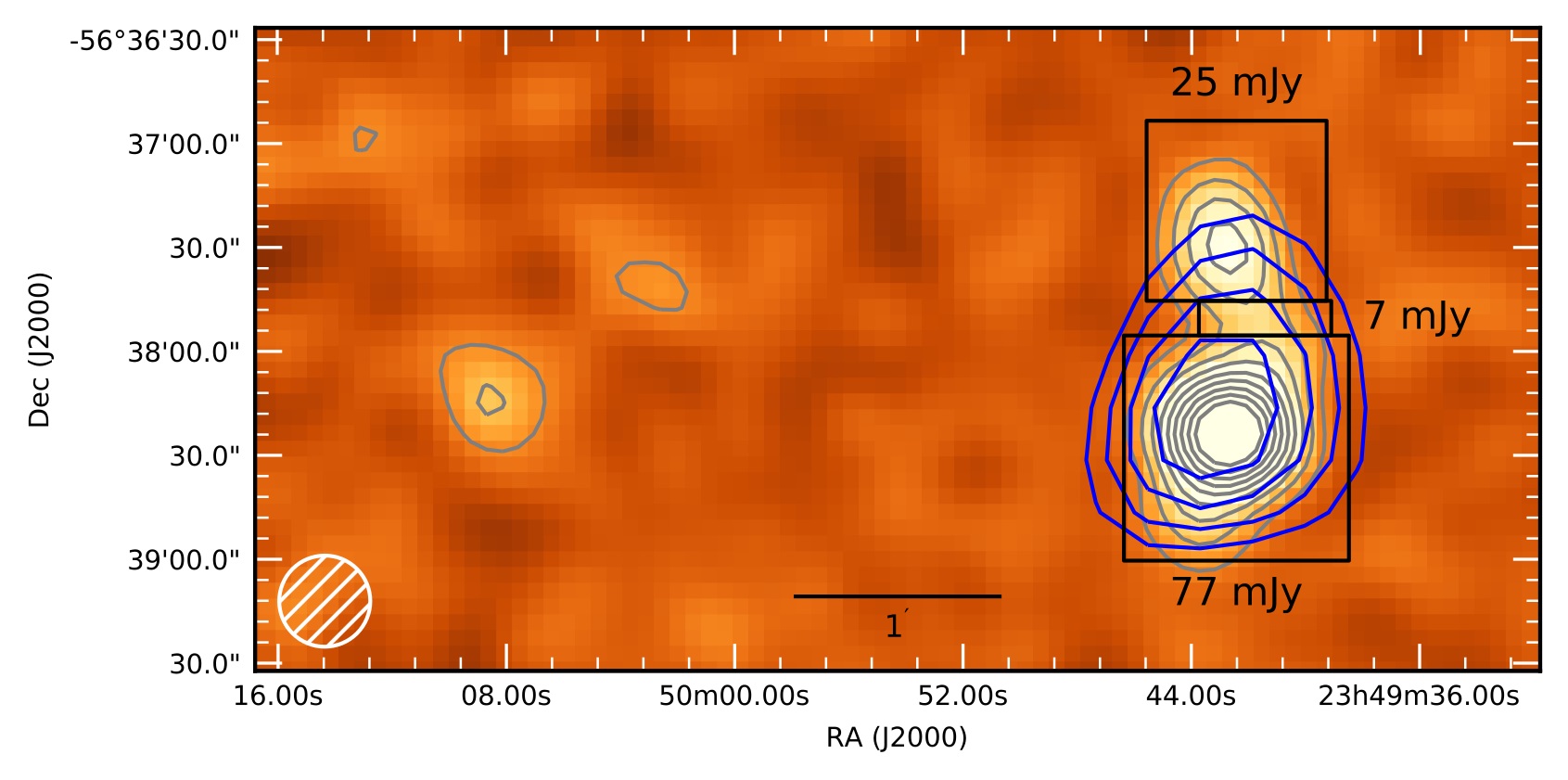}
	\caption{\textbf{Wide field 870-$\boldsymbol{\mu}$m image and photometry.}
		A wide-field LABOCA image (21$''$  beam size)
		of \spt. The image rms noise is 1.3\,mJy at center, increasing to 2mJy at the edges of the region shown. The total flux density recovered is 110.0$\pm$9.5\,mJy. Sub-apertures are drawn showing three different regions and their recovered flux densities. Grey contours start at 3$\sigma$ and increase in steps of 3$\sigma$.  SPT 1.4-mm contours are also shown (blue), revealing that even with the 1$'$ beam of SPT, \spt\ is  resolved. One additional submm source is detected at $>5\sigma$ in the LABOCA image to the east of the primary source, though {\it Herschel}-SPIRE photometry indicates that it is unlikely to be at $z\sim4$.}
	\label{fig:laboca}
\end{figure}

\begin{figure}
	\centering
	\includegraphics[width = 0.9\textwidth]{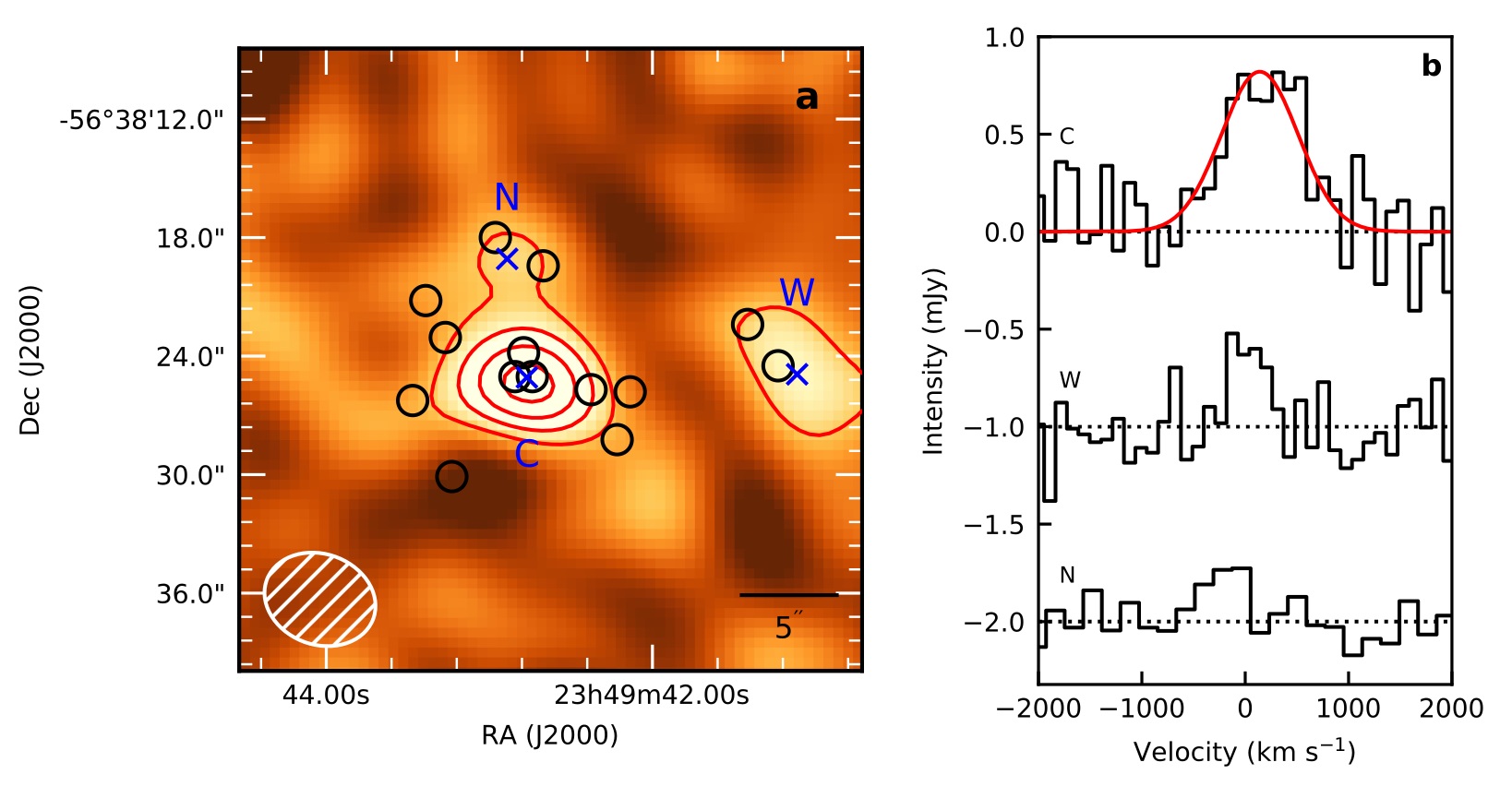}

	\caption{\textbf{CO(2-1) observations of \spt.} 
	(a):  The colormap and red contours trace the CO(2-1) line integrated over the central 830 \kms, with contours spaced by 2$\sigma$ starting at 2$\sigma$.
	The grey contours show the 1.1-mm ALMA continuum detections. 
	(b): One-dimensional spectra extracted at the positions indicated in (a).  	}
	\label{fig:atca}
\end{figure}

\begin{figure}
	\centering
	\includegraphics[width= 0.9\textwidth]{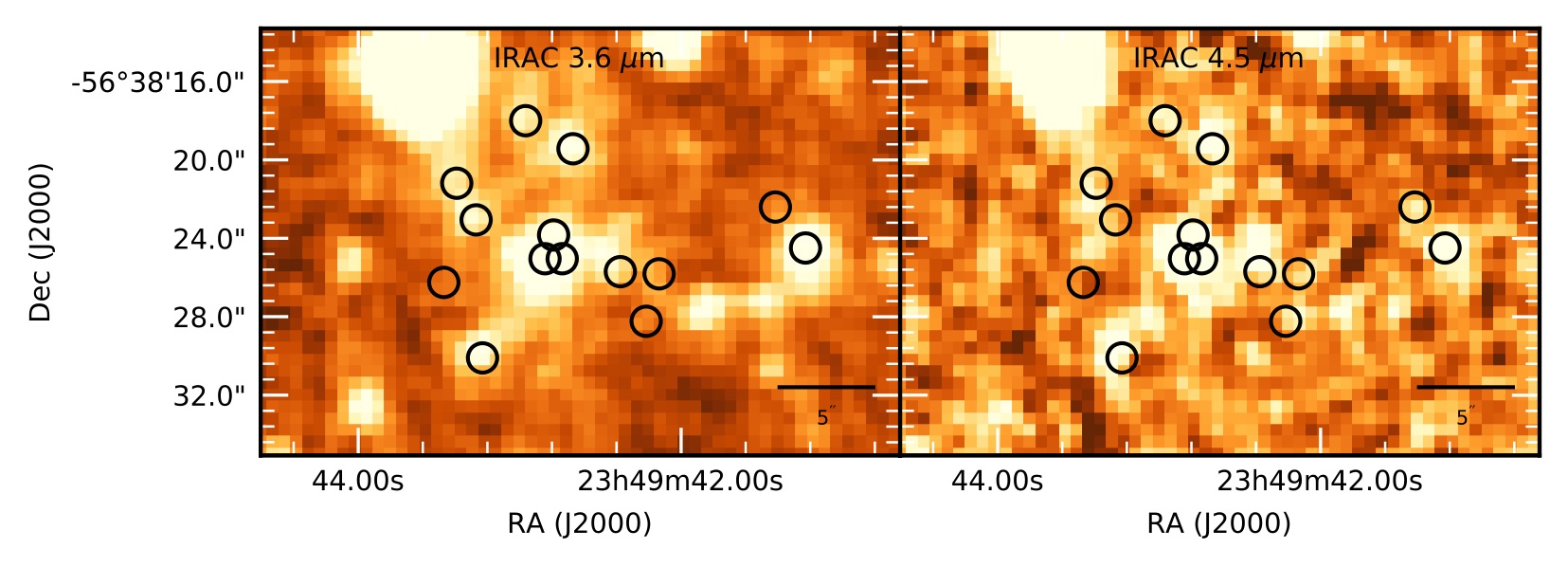}
	\caption{\textbf{IRAC observations of \spt.} Circles display the location of the 14 sources detected in ALMA band 7 described above. Nine of the 14 ALMA sources are detected in at least one of the IRAC bands with at least 3$\sigma$ confidence, including the two faintest [CII] sources (M \& N) from the blind line survey.  }
	\label{fig:IRAC}
\end{figure}

\begin{figure}
	\centering
	\includegraphics[width = 0.6\textwidth]{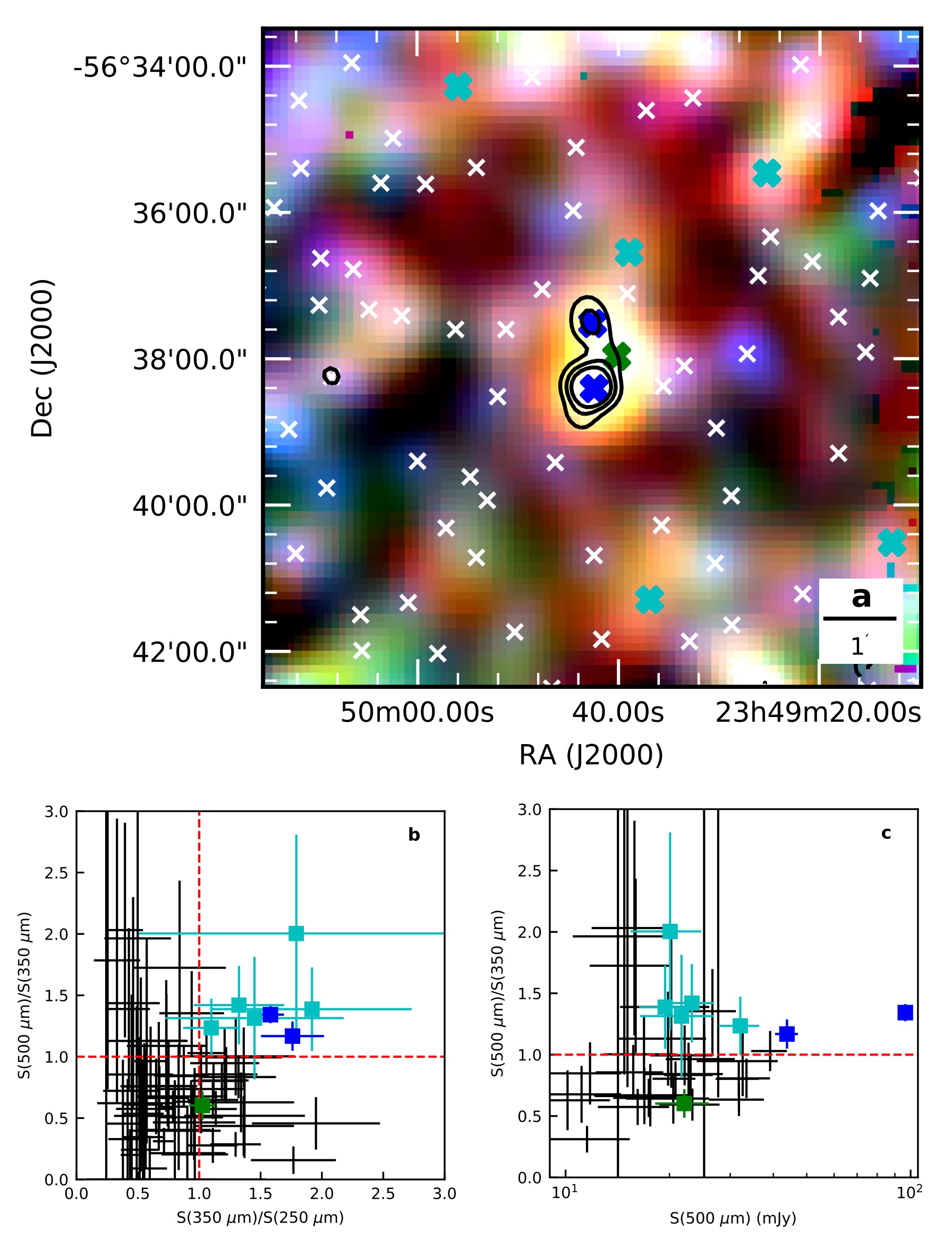}	
	\caption{\textbf{SPIRE RGB image and source colours in field surrounding \spt} (a) Deep SPIRE false colour image is shown with LABOCA contours overlaid.  Locations of 250-$\mu$m peaks used for analysis are marked with crosses (the faintest are not visible because of the contrast adopted in the image). Colour-colour (CC, (b)) and colour-flux (CF, (c)) diagrams for 250 $\mu$m sources are also shown. Error bars represent the 1 $\sigma$ standard deviation. The CC diagram shows sources with SNR(250\,$\mu$m)$>$3 and is dominated by the $z\sim1$ cosmic infrared background in the {\it foreground} of \spt\ (sources with colours ranging from blue to green). 
	The CF diagram applies an additional SNR(500\,$\mu$m)$>$3 cut. 
		The CC and CF diagrams show that one of three peaks associated with \spt\ is likely a lower-redshift interloper (green symbol), but also that there are five additional sources (blue symbols) in the surrounding region with colours ($S_{\rm 500\,\mu m} > S_{\rm 350\,\mu m} > S_{\rm 250\,\mu m}$) that are suggestive of $z$=4.3. 
	}
	\label{fig:spire}
\end{figure}

\begin{figure}
	\centering
	\includegraphics[width = 0.9\textwidth]{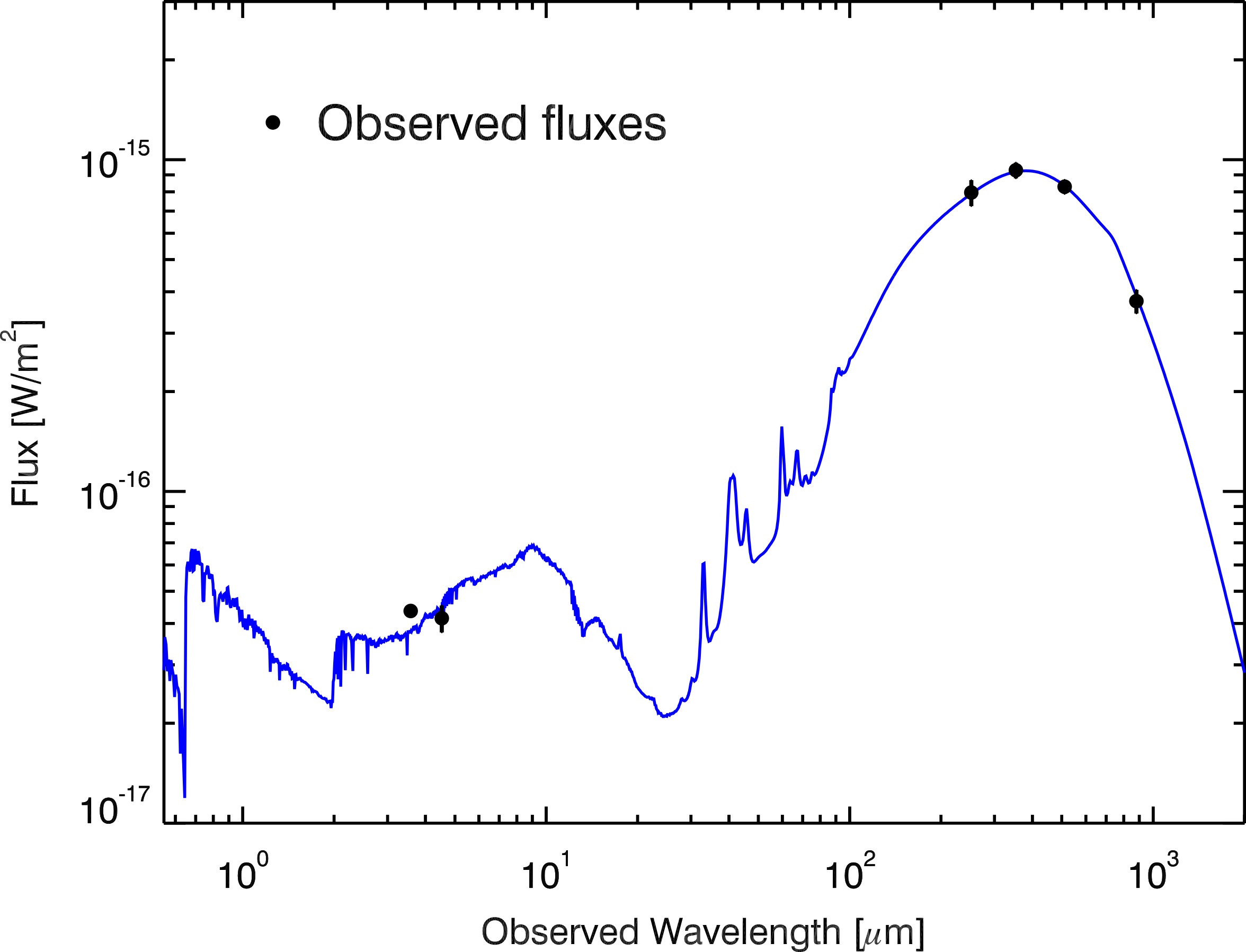}
	\caption{\textbf{Spectral energy distribution of \spt.} The SED of the extended \spt\ source is shown, including the summed deconvolved {\it Herschel}-SPIRE flux densities, the total 870-$\mu$m LABOCA flux density, and the summed IRAC flux densities.Error bars represent the 1 $\sigma$ measurement errors. We do not include the SPT 1.4, 2.0, and 3.0\,mm points because the source is elongated and flux measurements are difficult with the filtering used to make the map. Fitting the SED yields an IR luminosity of  $(8.0\pm1.0)\times10^{13}$ L$_\odot$.}
	\label{fig:SED}
\end{figure}

\begin{figure}
	\centering
	\includegraphics[width = 0.6\textwidth]{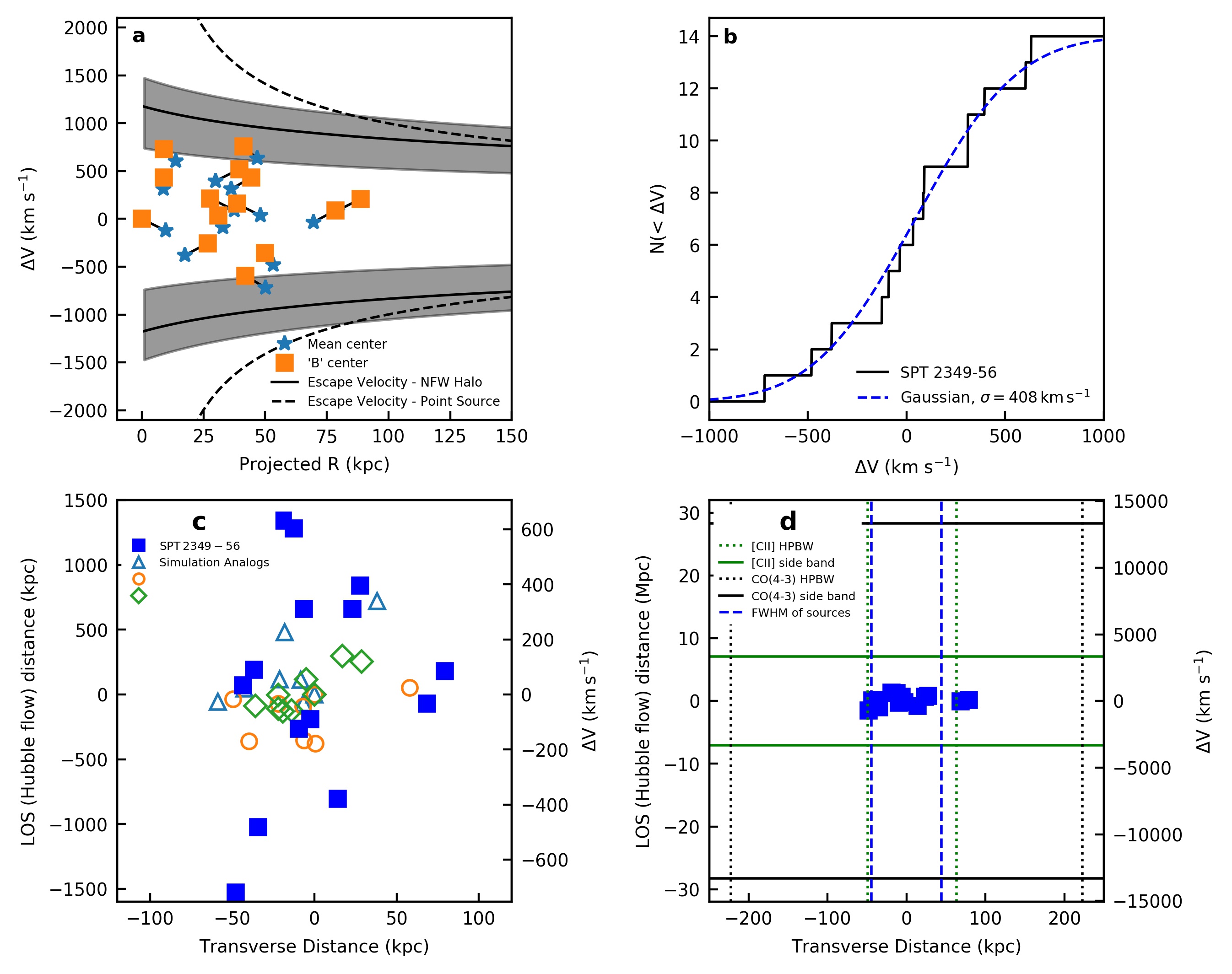}	
	\caption{\textbf{Geometry and dynamics of the \spt\ system.}
	(a) The velocity offsets versus projected (physical) distance are shown, compared to the escape velocity of a 1.16$\times$10$^{13}$ M$_\odot$ NFW halo (r$_{\rm vir}\sim$200 kpc and concentration $c = 5$), the grey shaded region showing our estimated halo mass uncertainty. Also shown is the escape velocity assuming a point mass halo of the same mass. Centres for the galaxy distribution are shown at the mean of the distribution, and centred on the `B' galaxy. All galaxies are bound for all but the lowest range of NFW halo masses.
	(b) The cumulative velocity distribution of the \spt\ galaxies, compared to a Gaussian distribution with our estimated dispersion (408 \kms), is at least consistent with expectations for a relatively bound system.
	(c) The physical distribution of the \spt\ galaxies is shown, assuming the redshifts are due to cosmic expansion rather than peculiar motions. This gives an extreme (but unlikely) possibility that \spt\ galaxies are stretched out along a {\it filament} compared to their 130kpc maximum tangential extent, but this requires that none of the velocity offsets are peculiar motions. The open markers show analogues of \spt\ found in a search specifically for maximally extended filamentary structures in the N-body simulations. These simulated structures that we have found are not filaments -- they are close to collapsed structures which are slightly cigar-shaped, and while \spt\ could in principle be distributed like this, it does not fundamentally change our discussion in the paper. 
We also note that  we are plotting 2 different things:
velocity offsets for \spt\ and actual geometry for the simulation galaxies, i.e. the 3-D positions. Even though the search allowed for structures extending $\sim 5$ Mpc along the line of sight, none are found to stretch beyond $1$ Mpc. 
	(d) Similar to panel (c) except the full extent of our ALMA band 3 \& 7 observations is shown. No structures are observed in the sidebands surrounding the 14 observed sources. }
	\label{fig:dv_dr}
\end{figure}
\clearpage
\bibliographystyle{naturemag}
\bibliography{library}
\end{document}